\documentclass[11pt,twoside,a4paper]{article}
\DeclareUnicodeCharacter{0229}{\c{e}}
\usepackage{graphicx}
\usepackage{dcolumn}
\usepackage{bm}

\usepackage[utf8]{inputenc}
\usepackage[T1]{fontenc}
\usepackage{mathptmx}

\usepackage{amssymb}
\usepackage{mathtools}
\usepackage{etoolbox}
\usepackage{braket}
\usepackage{tikz}
\usetikzlibrary{shapes.geometric,shapes.multipart, arrows}
\usepackage{bbold}
\usepackage{xcolor}
\usepackage{graphicx}
\graphicspath{ {./images/} }
\usepackage{caption}
\usepackage{subcaption}
\usepackage{tikzorbital}
\usepackage{tikz}\usepackage[sorting=none  , 
style=phys,
maxbibnames=99,
autocite=superscript]{biblatex} 
\usepackage{amsmath}
\usepackage{tikz-3dplot}
\usetikzlibrary{shapes.geometric,shapes.multipart, arrows, positioning, shapes,decorations,matrix,decorations.markings }
\tikzset{->-/.style={decoration={
  markings,
  mark=at position .5 with {\arrow{>}}},postaction={decorate}}}
\usepackage{float}
\usepackage{chemfig}
\usepackage{placeins}

\tikzstyle{startstop} = [rectangle, rounded corners, minimum width=6cm, minimum height=1cm,align=center, draw=black,inner sep=2pt,execute at begin node=]
\tikzstyle{process} = [rectangle, minimum width=6cm, minimum height=1cm, align=center, draw=black,inner sep=2pt,execute at begin node=]
\tikzstyle{decision} = [diamond, minimum width=6cm, aspect=2, align=center, draw=black,inner sep=0pt,execute at begin node=]
\tikzstyle{arrow} = [thick,->,>=stealth]

\newcommand{\onlinecite}[1]{\cite{#1}}

\newcommand{\Bohr}{a_0}

\usepackage{geometry}
\bibliography{ref}

\begin{document}

\title{The approximate Coupled-Cluster methods CC2 and CC3 in a finite magnetic field}
\author{Marios-Petros Kitsaras,$^{1,2}$ Laura Grazioli,$^2$ Stella Stopkowicz$^{1,2,3}$}
\date{$^1$Fachrichtung Chemie, Universit{\"a}t des Saarlandes, Campus B2.2, D-66123 Saarbr{\"u}cken, Germany \\
$^2$Department Chemie, Johannes Gutenberg-Universit\"at Mainz,
Duesbergweg 10-14, D-55128 Mainz, Germany \\
$^3$ Hylleraas Centre for Quantum Molecular Sciences, Department of Chemistry, University of Oslo, P.O. Box 1033 Blindern, N-0315 Oslo, Norway}

\maketitle

\begin{abstract}
In this paper, we report on the implementation of  CC2 and CC3 in the context of molecules in finite magnetic fields. 
The methods are applied to the investigation of atoms and molecules through spectroscopic predictions and  geometry optimizations for the study of the atmospheres of highly-magnetized White Dwarfs (WDs). 
We show that ground-state finite-field (ff) CC2 is a reasonable alternative to CCSD for energies and, in particular for geometrical properties. For excited states ff-CC2 is shown to perform well for states with predominant single-excitation character. 
Yet, for cases in which the excited-state wavefunction has double-excitation character with respect to the reference, ff-CC2 can easily
lead to completely unphysical results. 
Ff-CC3, however, is shown to reproduce the CCSDT behaviour very well and enables the treatment of larger systems at a high accuracy.

\end{abstract}

\section{Introduction}
In the recent decades, the chemistry in extreme environments has increased significantly as an active area of research. \cite{Schmelcher1997,M.R.Manaa2005,Fridman2008,Tellgren2009,Motzfeldt2012,Schmelcher2012,Lange2012,Miao2020,Margrave2022}
Studies of atoms and molecules in strong magnetic fields began already in the 80s and 90s\cite{Rosner1984,Schmelcher1988,Schmelcher1988a,Schmelcher1988b,Schmelcher1991,Jordan1998} and the interest  was revitalized in the 2000s for the study of molecular magnetic properties as well as molecules in magnetic fields with an arbitrary strength and orientation.\cite{Soncini2004,Tellgren2008,Tellgren2014,Sen2019,Sun2019b,Stopkowicz2015,Hampe2017,Furness2015,Reimann2019,Lehtola2020,Pausch2020,Pausch2021,Blaschke2021,Monzel2022}  The theoretical investigation of molecules in magnetic fields that are unattainable experimentally attracts special interest as it enables for example the study of the atmospheres of magnetic White Dwarfs stars (MWDs). The magnetic field on these celestial objects can be strong enough to compete with Coulomb interactions.\cite{Garstang1977,Jordan2008,Ferrario2015,Wickramasinghe2000,Deglinnocenti2004} The study of WDs aims to achieve a deeper understanding of the life cycle of stars and the evolution of the universe in general.\cite{Henry1984,Henry1985,Rosner1984,Forster1984,Greenstein1984,Greenstein1985,Schmidt1996,Schmidt1996err,Jordan1998,Dufour2007,Kawka2019,Berdyugina2007,Jordan2008,Jordan2008,Ferrario2015,Wickramasinghe2000,Deglinnocenti2004,Garstang1977} The additional fact that in the atmospheres of WD stars molecules are observed\cite{Berdyugina2007} offers a unique opportunity to study the exotic chemistry that arises when the magnetic interaction is at the same order of magnitude as the Coulomb interaction.\cite{Lange2012} 
Concerning quantum-chemical studies for such systems, both the magnetic as well as the Coulomb interaction need to be treated with non-perturbative quantum-chemical methods. 
These so-called finite-field (ff) methods require the use of complex algebra and deal with the gauge-origin dependence in the Hamiltonian. 
The use of complex algebra increases the computational cost, due to the need to handle complex rather than real values, and matrix multiplication is 3-4 times slower as compared to the real case.\cite{Dongarra1990}
Nowadays, gauge including atomic orbitals (GIAOs) also known as London orbitals\cite{London1937} are typically employed to ensure gauge-origin independent results.\cite{Tellgren2008} 
Implementations of different ff-methods in quantum-chemical programs that can handle arbitrarily oriented magnetic fields have been carried out in various program packages.\cite{QCUMBRE,cfour,LONDON,quest,bagel,WilliamsYoung2020,Pausch2020,Bischoff2020,Lehtola2020a}

Strong magnetic fields alter chemistry in such a way that  common chemical intuition is no longer applicable. 
The combination of the competing paramagnetic and diamagnetic influences may alter the ground state of atoms and molecules.\cite{Stopkowicz2015} Unusual phenomena 
like the perpendicular paramagnetic bonding,\cite{Lange2012} may also take place. Ff-Coupled-Cluster (CC) methods\cite{Cizek1966} are a valuable tool to study this complex behaviour, since they 
give highly accurate results with a clear way for systematic improvement.\cite{Stopkowicz2015} In addition, excited states may be studied by the ff-variants of Equation-of-Motion (EOM)-CC theory.\cite{Monkhorst1977,Stopkowicz2015,Hampe2017,Hampe2019,Hampe2020}

The most common CC variant used is probably "CC singles and doubles" (CCSD) theory. 
This method can be used for states described well by a single determinant and is applicable for small- to medium-sized molecules.\cite{Shavitt2009} Regarding larger molecules, its use is limited due to the unfavourable $N^6$ scaling, with $N$ the size of the system. In ff-calculations, the formal scaling is identical albeit with a larger prefactor. Molecular symmetry may be exploited using group theory to moderate the computational demands even in ff-calculations,\cite{Bishop1973,Davidson1975s,Taylor1986} though the symmetry is typically reduced by the magnetic field.\cite{Pausch2021s,unpKitsaras2022,Kitsaras2023}  Additionally, approaches that aim to lower the computational cost by approximating the four-center two-electron integrals like the Cholesky Decomposition (CD)\cite{Epifanovsky2013,Folkestad2019,Blaschke2021,Gauss2022}  or the Resolution of the Identity (RI)\cite{Reynolds2015,Pausch2020} have merit in combination with ff-CC methods, though they do not lower the overall scaling in CC calculations. Another way to moderate the cost of the standard CC truncations is by approximating the amplitude equations. Such an approximation is offered by the CC$n$ series of methods that lower the scaling of the parent method and are based on perturbation theory.\cite{Christiansen1995cc2,Christiansen1995cc3,Christiansen1995respcc3,Koch1997} 

CC2 is an approximation to CCSD and provides a lower scaling while often retaining a similar accuracy for energy differences.\cite{Christiansen1995cc2} The method is used exceedingly either to study larger molecules or to benchmark results from more approximate methods when CCSD is too expensive. \cite{Wiebeler2021,Stockett2020,Hornum2020,Dupuy2019,Durbeej2020,Yu2021,Reinholdt2021,Safin2021,Kochman2021,Naim2021,Izsak2020} The approximate CC2 method scales similarly to MP2, i.e., $N^5$, but, due to the inclusion of single excitations acting like an approximate orbital response, gives results of higher quality, especially for properties.\cite{Christiansen1995cc2,Hattig2006,Schreiber2008} Accordingly, spin contamination at the CC2 level is significantly reduced compared to MP2.\cite{Kitsaras2021} An essential feature of CC2 is that, unlike MP2, it can be rigorously extended to target excited states (EOM-CC2), thus enabling their treatment in larger systems at the CC level. 

In certain applications, an accuracy beyond CCSD may be needed. This is, for example, the case in studies of systems with moderate static-correlation character or excited states with a significant double-excitation character with respect to the reference.  
Since the next-higher-order truncation, CCSDT, scales as  $N^8$,  its applicability is  limited significantly.  For ground states, the  approximate triples scheme CCSD(T) is considered the "gold standard".\cite{Purvis1982} Only one non-iterative $N^7$ step is needed here, but the extension to excited states is not straight-forward. Moreover, the method is known to be sensitive to spin contamination.\cite{Krylov2017} The approximate inclusion of triple corrections via the CC3 model scales as $N^7$ as well, though iteratively, and is easily generalized to the EOM approach for treatment of excited states.\cite{Christiansen1995cc3,Koch1997} Recently, numerous applications of the CC3 approach have been reported: Highly accurate results at the CC3 level of theory are used for comparison with more approximate methods\cite{Bondanza2021} and interpretation of experimental data (UV spectroscopy, X-ray spectroscopy, organic photochemistry, etc.).\cite{Davis2021,Bilalbegovic2021,Hutcheson2021,Neville2018,Fedotov2021} CC3 has also contributed to the creation of databases and benchmarks for the electronic-structure community.\cite{Veril2021,Matthews2020a,Loos2020} Recent work towards more efficient implementations of CC3 for closed-shell systems by \citeauthor{Paul2021} has allowed applications to larger systems with more than 500 basis functions.\cite{Paul2021}
In the context of ff-methods, highly accurate results beyond CCSD are needed for example for the prediction and assignment of absorption spectra from MWD stars.\cite{Hampe2020} Additionally, when increasing the magnetic-field strength, double-excitation character can be transferred back and forth between various states.\cite{Hampe2020} The resulting problem of a deteriorating quality in the accuracy of the predictions can be dealt with by the inclusion of triple excitations, as shown in a study, at the ff-CCSDT level.\cite{Hampe2020} 
Hence, the ff-CC3 model, as is often the case for calculations in the absence of a magnetic field, may prove very useful to treat system where the full ff-CCSDT method is not feasible. 

In this paper, the implementation of the ff-CC2 and ff-CC3 methods 
is reported. It is based on a spin-unrestricted formulation that is able to target open-shell systems. Open-shell electronic configurations are typically stabilized in a magnetic field because of the spin-Zeeman influence and become ground states in stronger fields. 
In section \ref{sec:theory}, the theoretical aspects of the CC$n$ approximations are briefly presented as well as their extension to  EOM-CC theory.\cite{Christiansen1995cc2,Christiansen1995cc3,Koch1997} Details on the  implementation are discussed in section \ref{sec:implementation}. We note that ff-CC2 has been reported recently in the literature in terms of benchmark data,\cite{Pausch2021} but no dedicated implementation and investigation of its performance has been presented. Finally, applications of ff-CC2 and ff-CC3 are presented in section \ref{sec:results}. These include an investigation of the excitation spectrum of the Mg atom in the presence of a strong  magnetic field, calculations on the diatomic cation CH$^+$ and the CH radical, and geometry optimization in the presence of a magnetic field for methane CH$_4$ and ethylene CH$_2$CH$_2$.

\section{Theory \label{sec:theory}}
\subsection{Coupled Cluster Theory}
In the CC ansatz,\cite{Cizek1966} the wavefunction is written as the exponential of the cluster operator $\hat{T}$ acting on a reference wavefunction $\ket{0}$, which is usually the Hartree-Fock (HF) solution\cite{Shavitt2009}
\begin{equation*}
    \ket{\mathrm{CC}} = \mathrm{e}^{\hat{T}} \ket{0}.
\end{equation*}
Using the similarity transformed Hamiltonian
\begin{equation*}
    \widetilde{H} = \mathrm{e}^{-\hat{T}} \hat{H} \mathrm{e}^{\hat{T}},
\end{equation*}
the CC energy is given as:
\begin{equation*}
    E_{\mathrm{CC}} = \bra{0} \widetilde{H} \ket{0}.
\end{equation*}
The equations that determine the cluster amplitudes are the CC equations that require the projections on excited determinants $\bra{\mu_I}$ to vanish
\begin{equation*}
    0 = \bra{\mu_I} \widetilde{H} \ket{0}.
\end{equation*}

Standard CC approximations truncate the cluster operator to a specific excitation level, e.g. for CCSD: $\hat{T}=\hat{T}_1+ \hat{T}_2$, for CCSDT: $\hat{T}=\hat{T}_1+ \hat{T}_2+ \hat{T}_3$ etc. and the CC equations needed to determine the cluster amplitudes $t_I$ only make use of excited determinants up to the same excitation level. 

\subsection{The Equation of Motion approach}
In the EOM-CC approach, excited states are described by acting  with a linear excitation operator $\hat{R}$ on the CC wavefunction\cite{Monkhorst1977,Shavitt2009}
\begin{equation*}
    \ket{\mathrm{EOM}}=\hat{R}\ket{\mathrm{CC}}, 
\end{equation*}

The determination of the EOM amplitudes $r^k_I$ is achieved by solving the energy eigenvalue problem and results in the expression 

\begin{align}
    \bra{\nu_J}\left[\widetilde{H},\hat{R}^k\right]\ket{0} =&  \omega^k_{\mathrm{exc}} \bra{\nu_J}\hat{R}^k\ket{0} \nonumber \\
    \sum_I \bra{\nu_J}\left[\widetilde{H}, {\hat{\mu}_I}\right]\ket{0} r^k_I =&  \omega^k_{\mathrm{exc}}   r^k_J \label{eq:EOM_eig}. 
\end{align}
The EOM equation thus defines a CI-like eigenvalue problem, with $r^k_I$ being the $k$-th right-eigenvector solution, $\omega^k_{\mathrm{exc}}=E^k_{\mathrm{exc}}-E_{\mathrm{CC}}$ its eigenvalue, and ${\hat{\mu}_I}$ a string of quasi-particle creation operators.
Usually, the $\hat{R}$ operator is truncated at the same level as the CC truncation it is based on.
The   matrix elements $\bra{\nu_K}\left[\widetilde{H}, {\hat{\mu}_I}\right]\ket{0}$ form the connected contributions to the CC Jacobian matrix, which can be viewed as the gradient of the CC Lagrangian with respect to the $t_I$ amplitudes.\cite{Monkhorst1977}

\subsection{The CC$n$ approximation}
The CC$n$ series introduced by Christiansen \textit{et al.}\cite{Christiansen1995cc2,Christiansen1995cc3,Koch1997} aims to approximate the standard CC truncations with a more favourable scaling.  This series of approximations is based on the perturbation expansion of the CC energy and the M{\o}ller-Plesset (MP)\cite{Moller1934} partitioning of the Hamiltonian
\begin{align}
    \hat{H} &= \hat{F} + \hat{V}. \nonumber 
\end{align}
Here, the zeroth-order Hamiltonian corresponds to the sum of Fock operators $\hat{F}=\sum_a \hat{f}(a)$ and the fluctuation potential $\hat{V}=\hat{H}-\hat{F}$ is the perturbation. Unlike  many-body perturbation theory (MBPT),\cite{Shavitt2009} the CC$n$ series has two requirements: \cite{Christiansen1995cc2,Koch1997}
\begin{enumerate}
    \item The singles amplitudes $t_1$ are treated in zeroth order, as they function as effective orbital relaxation. \label{req:singles}
    \item The amplitude equations of excitation level $n$, where $n$ the cardinal number of the method, are simplified till the first non-vanishing order. \label{req:non-vanishing}
\end{enumerate} 

With the help of partially transformed operators
\begin{equation*}
    \hat{O}_1 = \mathrm{e}^{-\hat{T}_1} \hat{O} \mathrm{e}^{\hat{T}_1},
\end{equation*}
it is ensured that contributions of $\hat{T}_1$  will be included up to infinite order of perturbation. 

In the singles approximate doubles model, CC2,\cite{Christiansen1995cc2}  the double amplitude equations are truncated up to first order as dictated by the second requirement

\begin{align*}
    0=&\bra{\mu_2} 
    \hat{V}_1 +
    \left[\hat{F}_1,\hat{T}_2\right]  
      \ket{0}.
\end{align*}
Assuming the occupied-occupied and virtual-virtual blocks of the Fock matrix to be diagonal results in
\begin{align}
      0=&  \bra{\mu_2} \hat{V}_1\ket{0} + \Delta\epsilon_2 t_2   \nonumber \\
      t_2=&-\frac{\bra{\mu_2} \hat{V}_1\ket{0}}{\Delta\epsilon_2 }, \label{eq:CC2}
\end{align}
where $\Delta\epsilon_2$ signifies the orbital-energy differences $\epsilon_a+\epsilon_b-\epsilon_i-\epsilon_j$. In the notation used $a$,$b$,$c$,... signify virtual orbitals and $i$,$j$,$k$,... occupied orbitals.
For the singles doubles approximate triples model, CC3,\cite{Koch1997} one simplifies the triples amplitude equations till second order
\begin{align}
    0=&\bra{\mu_3} 
    \left[\hat{V}_1,\hat{T}_2\right] + 
    \left[\hat{F}_1,\hat{T}_3\right]  
    \ket{0} \nonumber \\
    t_3=&-\frac{\bra{\mu_3}\left[\hat{V}_1,\hat{T}_2\right]\ket{0}}{\Delta\epsilon_3}. \label{eq:CC3}
\end{align}
Again, canonical or semi-canonical orbitals are assumed. $\Delta\epsilon_3$ is the orbital-energy difference $\epsilon_a+\epsilon_b+\epsilon_c-\epsilon_i-\epsilon_j-\epsilon_k$. 

Building the CC$n$ Jacobian to form the EOM eigenvalue problem results in   
\begin{equation}
    \begin{pmatrix}
        \bra{\nu_1} \left[\hat{H}_1,\hat{\mu}_1\right] + \left[\left[\hat{H}_1,\hat{\mu}_1\right],\hat{T}_2\right] \ket{0} &  \bra{\nu_1} \left[\hat{H}_1,\hat{\mu}_2\right] \ket{0}  \\
        \bra{\nu_2}\left[\hat{V}_1,\hat{\mu}_1\right] \ket{0}  & \bra{\nu_2}  \left[\hat{F},\hat{\mu}_2\right] \ket{0} 
    \end{pmatrix}
    \begin{pmatrix}
        r_1 \\
        r_2
    \end{pmatrix}
    =
    \omega_{\mathrm{exc}}
    \begin{pmatrix}
        r_1 \\
        r_2
    \end{pmatrix} \label{eq:EOM_CC2}
\end{equation}
for EOM-CC2\cite{Christiansen1995cc2} and 
\begin{equation}
\resizebox{.93\hsize}{!}{$
    \begin{pmatrix}
        \bra{\nu_1} \left[\hat{H}_1,\hat{\mu}_1\right] +
    \left[\left[\hat{H}_1,\hat{\mu}_1\right],\hat{T}_2\right]
    \ket{0} & 
    \bra{\nu_1} 
    \left[\hat{H}_1,\hat{\mu}_2\right] 
    \ket{0} &  
    \bra{\nu_1}
    \left[\hat{H},\hat{\mu}_3\right]  
    \ket{0} \\ 
    \begin{aligned}
        \bra{\nu_2}  \left[\hat{H}_1,\hat{\mu}_1\right] +
        \left[\left[\hat{H}_1,\hat{\mu}_1\right],\hat{T}_2 + \hat{T}_3\right]  \ket{0}
    \end{aligned}
    & 
    \begin{aligned}
        \bra{\nu_2}  \left[\hat{H}_1,\hat{\mu}_2\right] + 
        \left[\left[\hat{H}_1,\hat{T}_2\right],\hat{\mu}_2\right]   \ket{0} 
    \end{aligned}
    & \bra{\nu_2}  
    \left[\hat{H}_1,\hat{\mu}_3\right]  \ket{0} \\ 
    \bra{\nu_3}\left[\left[\hat{V}_1,\hat{\mu}_1\right],\hat{T}_2\right]\ket{0}  & \bra{\nu_3}\left[\hat{V}_1,\hat{\mu}_2\right] 
    \ket{0} & \bra{\nu_3}
    \left[\hat{F},\hat{\mu}_3\right]  
    \ket{0}
    \end{pmatrix}
    \begin{pmatrix}
        r_1 \\ 
        r_2 \\ 
        r_3
    \end{pmatrix} \\
    =
    \omega_{\mathrm{exc}}
    \begin{pmatrix}
        r_1 \\
        r_2 \\
        r_3
    \end{pmatrix} \label{eq:EOM_CC3}
    $
    }
\end{equation}
\normalsize
for EOM-CC3.\cite{Christiansen1995cc3,Paul2021}

Noting that within EOM-CC2, double-excitation amplitudes $r_2$ are fully determined by the single amplitudes $r_1$ (see also sec.~\ref{subsec:cons}), one could rewrite the EOM-CC2 eigenvalue problem as a non-linear set of equations involving only $r_1$ amplitudes. For those cases in which the double amplitudes are the \textit{leading} contributions for the excited state, the approximation breaks down, meaning that states with a predominant double-excitation character  cannot be targeted by the EOM-CC2 approach.\cite{Christiansen1995cc2} The same is true for CC3 and excited states with predominant triple-excitation character, but such states are of no particular concern in practical applications.

\section{Implementation \label{sec:implementation}}

The implementation of ff-CC2 and ff-CC3 as well as the EOM approach of these methods has been carried out in the QCUMBRE program package.\cite{QCUMBRE} 
\subsection{General Considerations \label{subsec:cons}}
The scaling of the (EOM-)CC2 and (EOM-)CC3 methods is $N^5$ and $N^7$ respectively, which is one order of magnitude less than their parent methods  CCSD and CCSDT. 
Beyond this reduction of the computational cost, an efficient implementation can be achieved, when using canonical or semi-canonical orbitals. The diagonal form of the Fock matrix results in a diagonal form of the $n$-th amplitudes equations.  Hence, double amplitudes can be expressed as a function of the single amplitudes alone with no doubles to doubles contributions for CC2 (see eq. \ref{eq:CC2}). Respectively for CC3, the triple amplitudes equations can be brought into a form that depends only on the singles and doubles with no triples to triples contributions (see eq.~\ref{eq:CC3}). 
This commonly used approach allows for an \textit{on-the-fly} calculation of the amplitudes which reduces the memory requirements.\cite{Wuellen2016}

Turning to the Jacobian matrices, similar equations have been derived for the approximated $r_n$ amplitudes for excited states, where $n$ is the cardinal number of the method.
Starting from the last row of eq. (\ref{eq:EOM_CC2}), the expression 
\begin{align}
    \bra{\nu_2} \left[\hat{V}_1 , \hat{\mu}_1 \right] \ket{0} r_1 + \Delta\epsilon_2 r_2 =&  \omega_{\mathrm{exc}} r_2 \nonumber \\
    \frac{\bra{\nu_2} \left[\hat{V}_1 , \hat{\mu}_1 \right] \ket{0} r_1}{\omega_{\mathrm{exc}}-\Delta\epsilon_2}  =&   r_2  \label{eq:CC2_r2}
\end{align}
is obtained for CC2. Similarly, continuing from the last row of eq. (\ref{eq:EOM_CC3}) leads to the diagonal $r_3$ elements of EOM-CC3
\begin{align}
    \bra{\nu_3}\left[\left[\hat{V}_1,\hat{\mu}_1\right],\hat{T}_2\right]\ket{0} r_1 + \bra{\nu_3}\left[\hat{V}_1,\hat{\mu}_2\right] 
    \ket{0} r_2 + \Delta\epsilon_3 r_3 =&  \omega_{\mathrm{exc}} r_3 \nonumber \\
    \frac{\bra{\nu_3}\left[\left[\hat{V}_1,\hat{\mu}_1\right],\hat{T}_2\right]\ket{0} r_1 + \bra{\nu_3}\left[\hat{V}_1,\hat{\mu}_2\right] 
    \ket{0} r_2}{\omega_{\mathrm{exc}}-\Delta\epsilon_3}  =&   r_3 . \label{eq:CC3_r3}
\end{align}

Eq. (\ref{eq:CC2_r2}) and (\ref{eq:CC3_r3}) show, similarly to the ground state treatment, that the $r_n$ amplitudes can be fully determined by the lower excitation levels and can thus be considered redundant information not to be saved in memory. Unlike the $t_n$ amplitudes however, they require the calculation of the exact excitation energy, which is  available only at convergence. It is well known and exploited in many field-free EOM-CC$n$ implementations that to deal with this issue, the Davidson method in EOM-CC$n$ calculations needs to be modified. Details on these can be found in ref.~\onlinecite{Kitsaras2023}.

\subsection{Validation}
 To verify the implementation, the code has been tested for the field-free case against calculations using the closed-shell implementation of the CFOUR program.\cite{cfour,Matthews2020}  For calculations in a finite magnetic field and for the case of an unrestriced reference, the CCSD and CCSDT methods already implemented in QCUMBRE were modified to produce results at the (EOM-)CC2 and CC3 levels respectively to verify the more efficient CC2 and CC3 implementations.

\section{Results and Discussion \label{sec:results}}
All post-HF calculations were performed using the QCUMBRE program package.\cite{QCUMBRE} QCUMBRE works together with an interface to the CFOUR program\cite{cfour,Matthews2020} that provides integrals over London orbitals via the  MINT integral code\cite{MINT} and a ff-UHF reference wavefunction.

First, the investigation of the Mg atom will be presented, followed by  calculations on the CH$^+$ and CH molecules. Lastly, the results of geometry optimizations for methane and ethylene in different magnetic fields will be presented. The discussion in the following paragraphs is centered on the performance of the CC2 and CC3 methods, while an investigation of the physical and chemical behaviour of the systems in the presence of a magnetic field is presented in ref.~\onlinecite{Kitsaras2023}.

\subsection{Atomic Mg}
In the atmospheres of WDs, metals are often detected due to incoming material from planetary or asteroidal debris.\cite{Kawka2019,Zuckerman2003,Zuckerman2010} 
In the current study, highly accurate results for the Mg atom have been generated at the CC3 level of theory in order to investigate such contaminants in magnetic WDs. Specifically, transitions from the lowest triplet state $^3P_u$ described by electronic configuration $1s^22s^22p^63s^13p^1$ to triplet states $^3S_g$ ($1s^22s^22p^63s^14s^1$) and $^3D_g$ ($1s^22s^22p^63s^13d^1$) were investigated as they are expected to give strong signals.\cite{nist}  The electronic states involved were studied at the CCSD and CC3 levels of theory using a series of uncontracted (unc) basis sets, namely the unc-aug-cc-pCV$X$Z sets,\cite{pritchard2019a,feller1996a,schuchardt2007a,prascher2011a} with $X$ the cardinal number of the basis set. Using the  \mbox{unc-aug-cc-pCVTZ} basis, additional calculations at the CCSDT level of theory were performed to benchmark the CC3 results. The magnitude of the triples corrections was calculated as the difference between the CCSDT and CCSD energy. 
Calculations were performed in the range of field strengths between $0.0\ B_0$ and $0.2\ B_0$. A dense spacing of $0.004\ B_0$ was used up to $0.1 \ B_0$, continuing with a spacing of $0.02\ B_0$ for the last increment.  

\begin{figure}[ht]
\begin{minipage}[t]{0.47\textwidth}

\begin{tikzpicture}
    \node[anchor=south west,inner sep=0] (image) at (0,0) {\includegraphics[width=\textwidth]{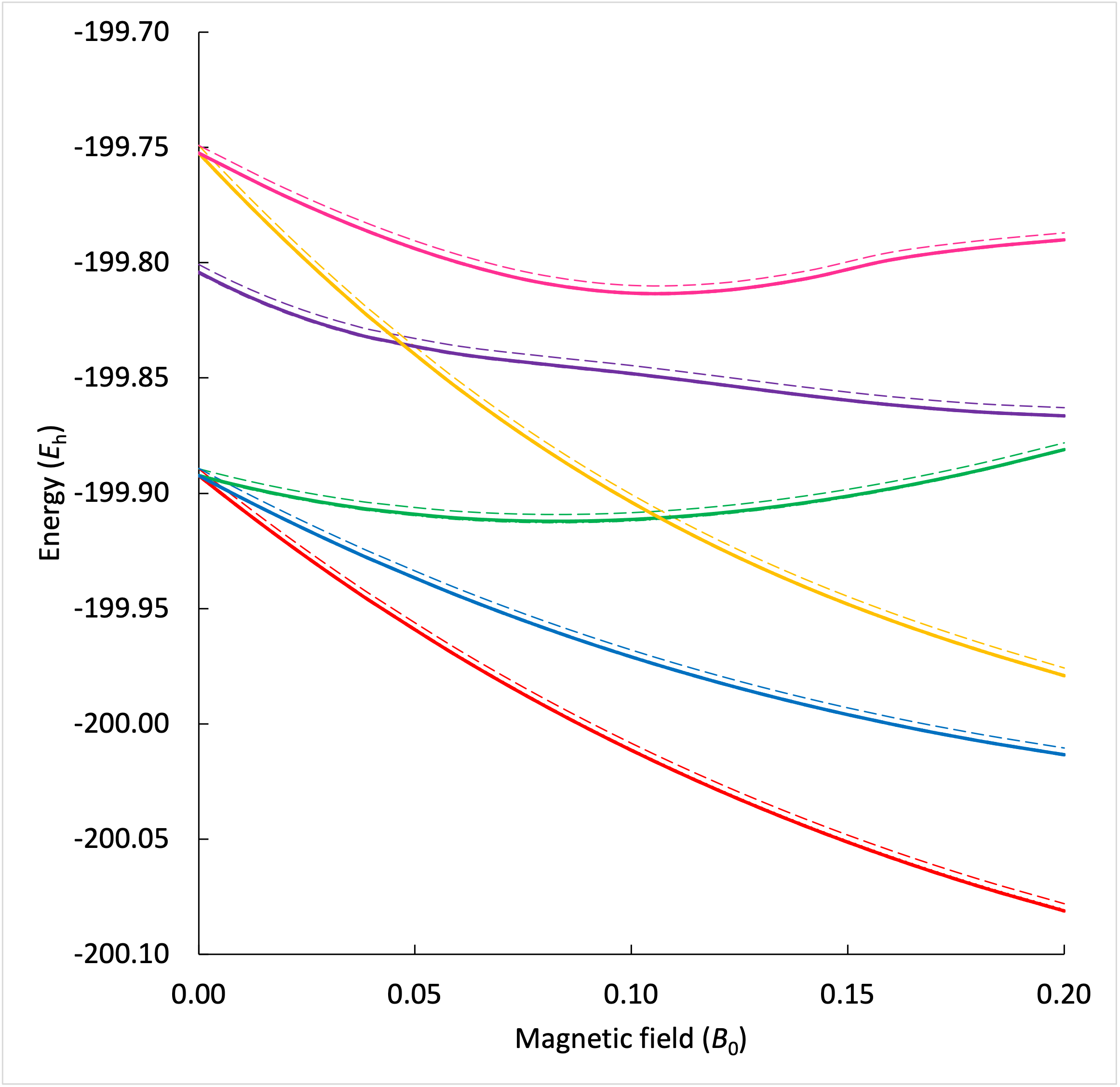}};
    \begin{scope}[x={(image.south east)},y={(image.north west)}]
        \node at (0.78,0.16) (label1) {\scalebox{0.75}{$^3\Pi_u^- ({^3P}_u)$}};
        \node at (0.7,0.32) (label2) {\scalebox{0.75}{$^3\Sigma_u ({^3P}_u)$}};
        \node at (0.55,0.48) (label3) {\scalebox{0.75}{$^3\Pi_u^+ ({^3P}_u)$}};
        \node at (0.84,0.46) (label4) {\scalebox{0.75}{$^3\Delta_g^- ({^3D}_g)$}};
        \node at (0.83,0.67) (label5) {\scalebox{0.75}{$^3\Sigma_g ({^3S}_g)$}};
        \node at (0.84,0.81) (label6) {\scalebox{0.75}{$^3\Sigma_g ({^3D}_g)$}};
        \draw[->] (0.2,0.57) -- (0.2,0.72);
        \draw[->] (0.22,0.57) -- (0.22,0.83);

        \node[anchor=west,minimum height = 0.5 cm] at (0.25,0.95)  {\scalebox{0.8}{CCSD}};
        \draw[loosely dashed] (0.25,0.95) -- +(-0.06,0) node [midway] (L) {};
         \node[anchor=west,minimum height = 0.5 cm] at (0.25,0.9)  {\scalebox{0.8}{CC3}};
        \draw[densely dashed] (0.25,0.9) -- +(-0.06,0) node [midway] (L) {};
        \node[anchor=west,minimum height = 0.5 cm] at (0.5,0.95)  {\scalebox{0.8}{CCSDT}};
        \draw[-] (0.5,0.95) -- +(-0.06,0);
    \end{scope}
\end{tikzpicture}
\caption{Low-lying triplet states of Mg calculated at the EOM-CCSD, EOM-CC3, and EOM-CCSDT levels of theory using an unc-aug-cc-pCVTZ basis set. }
\label{fig:Mg_sp1}

\end{minipage} \hspace{0.02\textwidth}
\begin{minipage}[t]{0.47\textwidth}

\begin{tikzpicture}
    \node[anchor=south west,inner sep=0] (image) at (0,0) {\includegraphics[width=\textwidth]{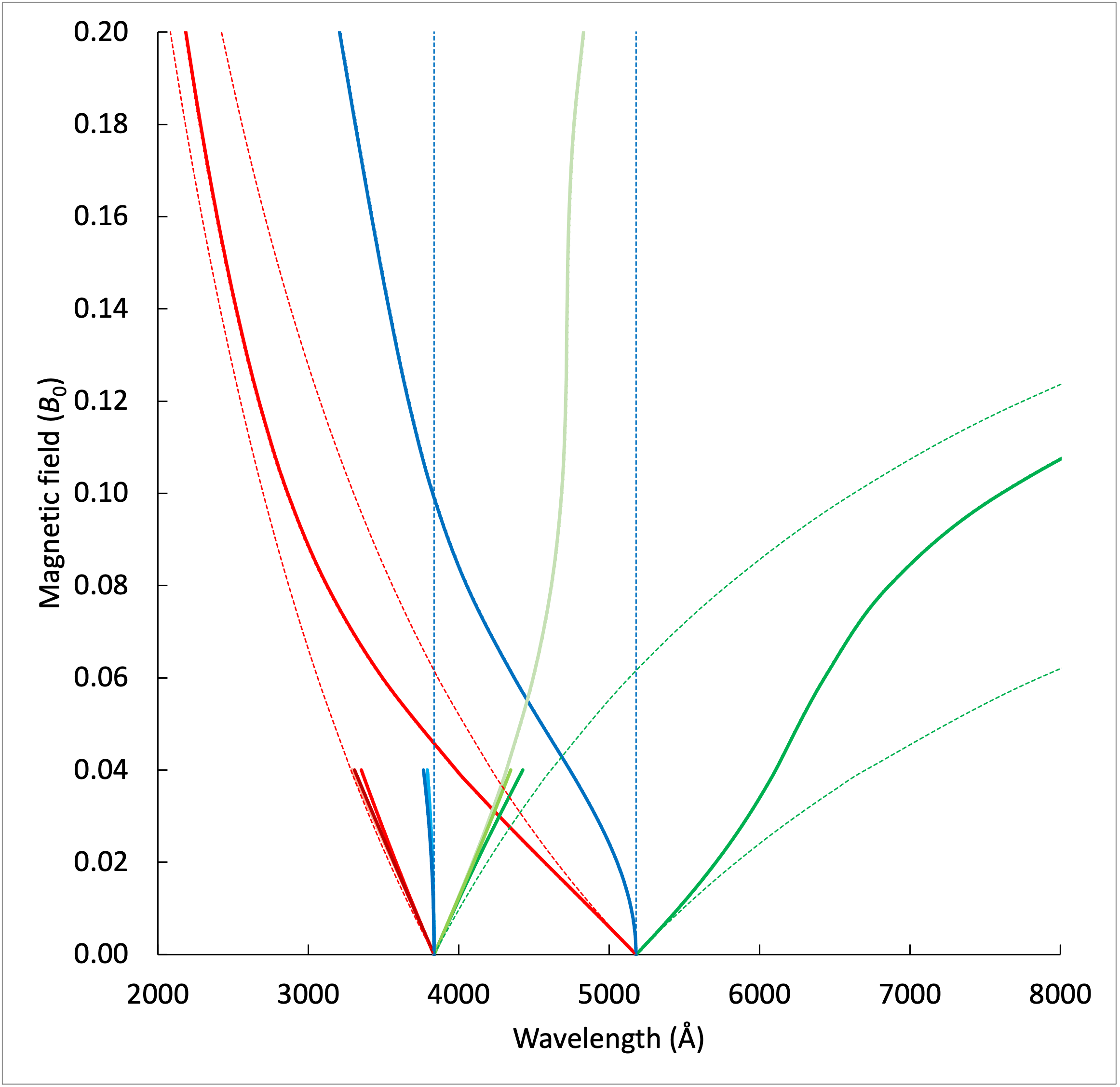}};
    \begin{scope}[x={(image.south east)},y={(image.north west)}]
        \node at (0.7,0.15) (label1) {\scalebox{0.75}{$^3P_u  \rightarrow {^3S}_g$}};
        \node at (0.28,0.15) (label2) {\scalebox{0.75}{$^3P_u  \rightarrow {^3D}_g$}};
        \node[anchor=west,minimum height = 0.5 cm] at (0.7,0.95)  {\scalebox{0.8}{Zeeman splitting}};
        \draw[dotted] (0.7,0.95) -- +(-0.06,0) node [midway] (L) {};
        \node[anchor=west,minimum height = 0.5 cm] at (0.7,0.9)  {\scalebox{0.8}{prediction}};
        \draw[-] (0.7,0.9) -- +(-0.06,0);
    \end{scope}
\end{tikzpicture}
\caption{The extrapolated B-$\lambda$ curves for the $^3P_u  \rightarrow {^3S}_g$ and $^3P_u  \rightarrow {^3D}_g$ transitions of Mg. The $\Delta M_L=0$ components are shown in  blue, the $\Delta M_L=+1$ in  red and the $\Delta M_L=-1$ in  green. The $^3P_u  \rightarrow {^3S}_g$ transition and one component of the  $^3P_u  \rightarrow {^3D}_g$ have been studied up to $0.20\ B_0$, while the rest of the transitions has been studied  up to $0.04\ B_0$.  }
\label{fig:Mg_bl}
\end{minipage}
    \end{figure}

The results for the total energy at the various levels of theory are shown in fig.~\ref{fig:Mg_sp1}. Following a similar strategy as in ref. \onlinecite{Hampe2020}, an extrapolation scheme has been employed to generate accurate B-$\lambda$ curves. In these curves, the magnetic-field strength is plotted as a function of the transition wavelength in fig.~\ref{fig:Mg_bl}.
The extrapolation scheme used for the generation of accurate B-$\lambda$ curves follows 
\begin{equation}
    \Delta E_{\mathrm{exc}}^{\text{corrected}}=\Delta E_{\mathrm{exc}}+\Delta E_{\mathrm{basis}}+\Delta E_{\text{triples}}.
\end{equation}
For the extrapolation, the CCSD/unc-aug-cc-pV$5Z$ results  were used for the excitation energies $\Delta E_{\mathrm{exc}}$ and the unc-aug-cc-pCVQZ and unc-aug-cc-pCV$5$Z basis sets for the basis-set extrapolation $\Delta E_{\mathrm{basis}}$.\cite{Hampe2020} Higher-order correlation was accounted for via triples corrections $\Delta E_{\text {triples}} = E_{\text {triples }}-E_{\text {CCSD}}$ at the  CC3/unc-aug-cc-pCVQZ levels of theory. Lastly, an offset correction relative to the NIST spin-averaged reference\cite{nist} is added.  The selection rule $\Delta M_L=0,\pm1$ ($0$ blue, $+1$ red, $-1$ green) was used to construct visible transitions.  In the figure, the results of a simple perturbative Zeeman  correction $E_\mathrm{Zeeman} = \frac{1}{2}M_L B $ were plotted as well (dotted curves). The deviation from this simple correction shows the importance of ff-quantum-chemical predictions for the assignment of spectra at  high-field strengths.

As seen in fig.~\ref{fig:Mg_sp1}, the CCSD energies  (long dashed curves) are energetically slightly higher as compared to the CCSDT reference. In contrast, the CC3 results are practically identical to CCSDT and cannot be distinguished in the plot. The calculated deviation of the approximate triples relative to the inclusion of full triples is at the order of only $10^{-5}\ E_\mathrm{h}$. The lower computational cost of CC3 compared to CCSDT and the good agreement between the two methods enables an accurate treatment of triple excitations using larger basis sets that is not feasible with a full treatment of triples.

It is important to note, that the use of the CC3 method for the generation of the extrapolated \mbox{B-$\lambda$} curves has allowed the use of triple corrections  calculated with larger basis sets. The comparison with the CCSDT results reveals the potential of CC3 to practically replicate the full inclusion of triples corrections for the electronic transitions studied. Moreover, the developments presented here contributed to the  assignment of Mg in the spectrum of a magnetised WD,\cite{Hollands2023} which would have not been possible without ff-quantum-chemical predictions. 

\FloatBarrier

\subsection{CH$^+$ and CH in varying magnetic fields \label{sec:CH_CH+}}
Calculations on the closed-shell methylidinium cation CH$^+$ and the open-shell CH radical have been carried out using a contracted and an  unc-cc-pVDZ basis set, respectively. The CH radical has been detected in weakly magnetised WDs along with C$_2$.\cite{Berdyugina2007,Jordan2008} This is an indication that the radical or its cation may be present in strongly magnetic WDs as well.

\subsubsection{The methylidinium cation CH$^+$}
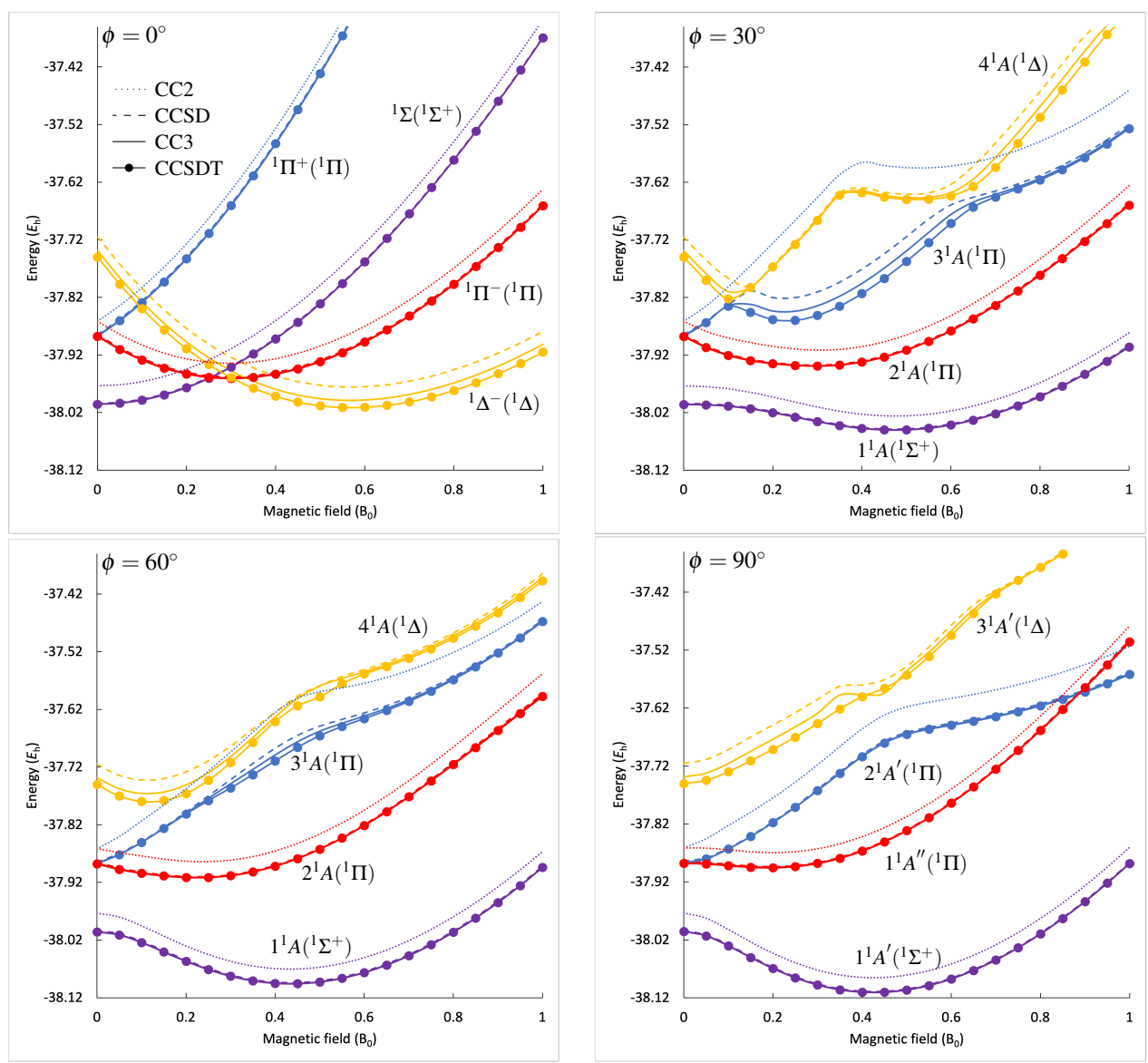
\begin{figure}[b!]
\centering
\begin{minipage}{\textwidth}
    \begin{minipage}[t]{0.47\textwidth}
    \begin{tikzpicture}
        \node[anchor=south west,inner sep=0] (image) at (0,0) {\includegraphics[width=\textwidth]{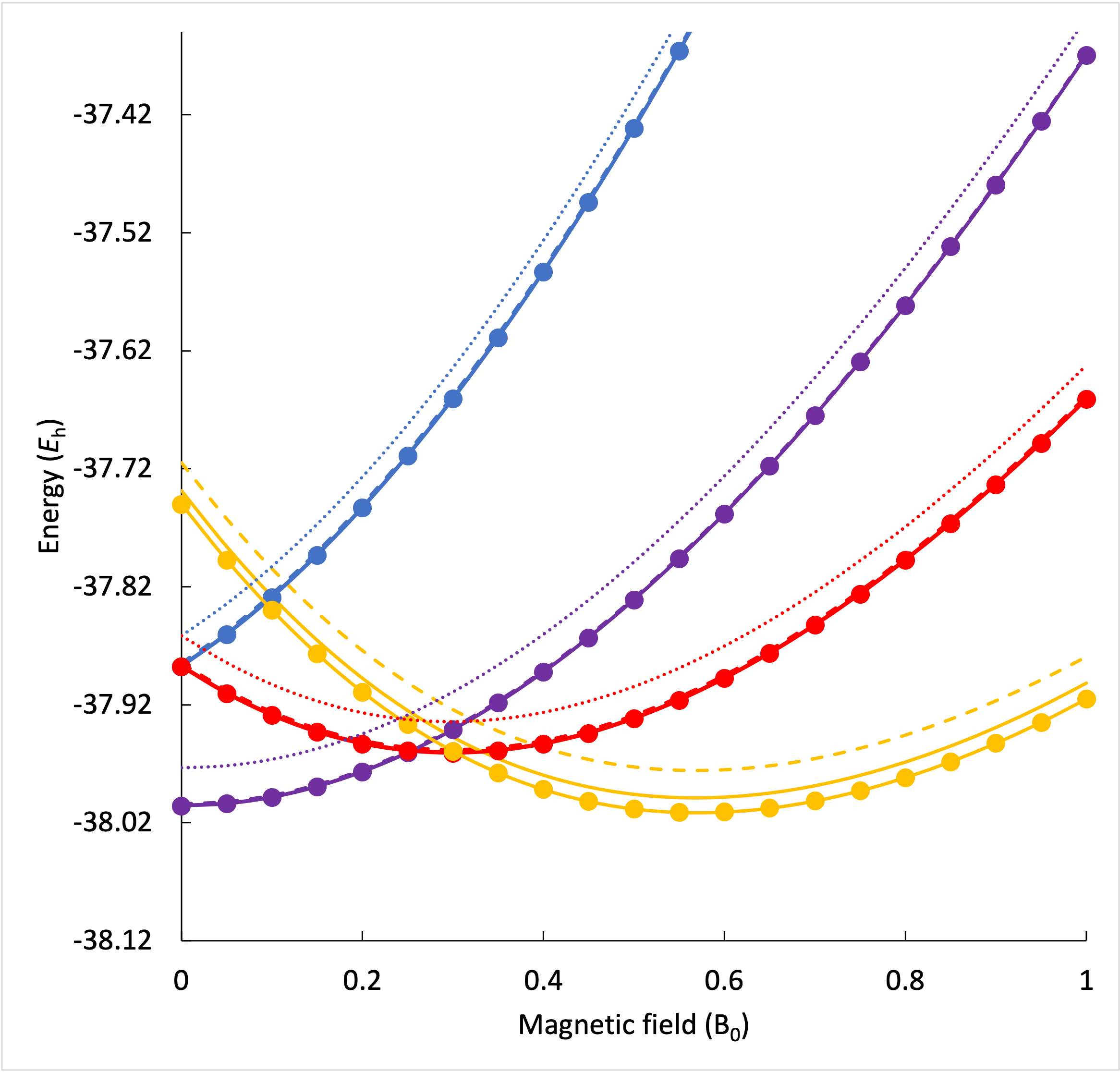}};
        \begin{scope}[x={(image.south east)},y={(image.north west)}]
            \node at (0.23,0.95) (angle) {\scalebox{0.9}{$\phi=0^\circ$}};
            \node at (0.76,0.8) (label1) {\scalebox{0.8}{$^1\Sigma (^1\Sigma^+)$}};
            \node at (0.9,0.46) (label2) {\scalebox{0.8}{$^1\Pi^- (^1\Pi)$}};
            \node at (0.55,0.7) (label6) {\scalebox{0.8}{$^1\Pi^+ (^1\Pi)$}};
            \node at (0.9,0.25) (label6) {\scalebox{0.8}{$^1\Delta^- (^1\Delta)$}};
            \node[anchor=west,minimum height = 0.5 cm] at (0.25,0.85)  {\scalebox{0.8}{CC2}};
            \draw[dotted] (0.25,0.85) -- +(-0.06,0) node [midway] (L) {};
            \node[anchor=west,minimum height = 0.5 cm] at (0.25,0.8)  {\scalebox{0.8}{CCSD}};
            \draw[dashed] (0.25,0.8) -- +(-0.06,0) node [midway] (L) {};
            \node[anchor=west,minimum height = 0.5 cm] at (0.25,0.75)  {\scalebox{0.8}{CC3}};
            \draw[-] (0.25,0.75) -- +(-0.06,0) node [midway] (L) {};
            \node[anchor=west,minimum height = 0.5 cm] at (0.25,0.7)  {\scalebox{0.8}{CCSDT}};
            \draw[-] (0.25,0.7) -- +(-0.06,0) node [midway] (L) {};
            \filldraw[black] (L)  circle  (1.5pt);

        \end{scope}
    \end{tikzpicture}
    \end{minipage} \hspace{0.02\textwidth}
    \begin{minipage}[t]{0.47\textwidth}
    \begin{tikzpicture}
        \node[anchor=south west,inner sep=0] (image) at (0,0) {\includegraphics[width=\textwidth]{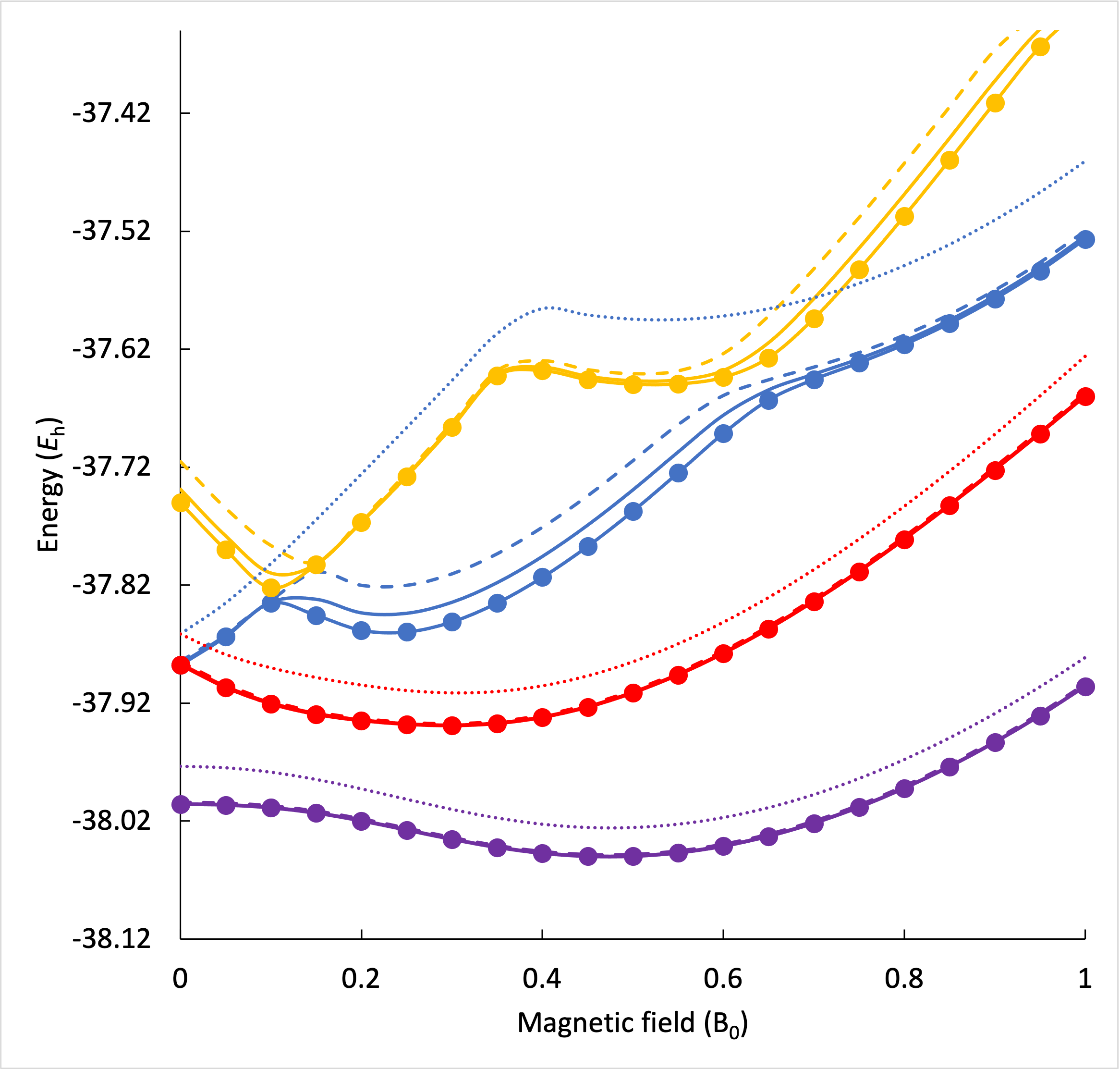}};
        \begin{scope}[x={(image.south east)},y={(image.north west)}]
            \node at (0.24,0.95) (angle) {\scalebox{0.9}{$\phi=30^\circ$}};
            \node at (0.55,0.16) (label1) {\scalebox{0.8}{$1{^1A} (^1\Sigma^+)$}};
            \node at (0.6,0.31) (label2) {\scalebox{0.8}{$2{^1A} (^1\Pi)$}};
            \node at (0.68,0.53) (label6) {\scalebox{0.8}{$3{^1A} (^1\Pi)$}};
            \node at (0.76,0.9) (label6) {\scalebox{0.8}{$4{^1A} (^1\Delta)$}};

        \end{scope}
    \end{tikzpicture}
    \end{minipage}
\end{minipage}

\begin{minipage}{\textwidth}
    \begin{minipage}[t]{0.47\textwidth}
    \begin{tikzpicture}
        \node[anchor=south west,inner sep=0] (image) at (0,0) {\includegraphics[width=\textwidth]{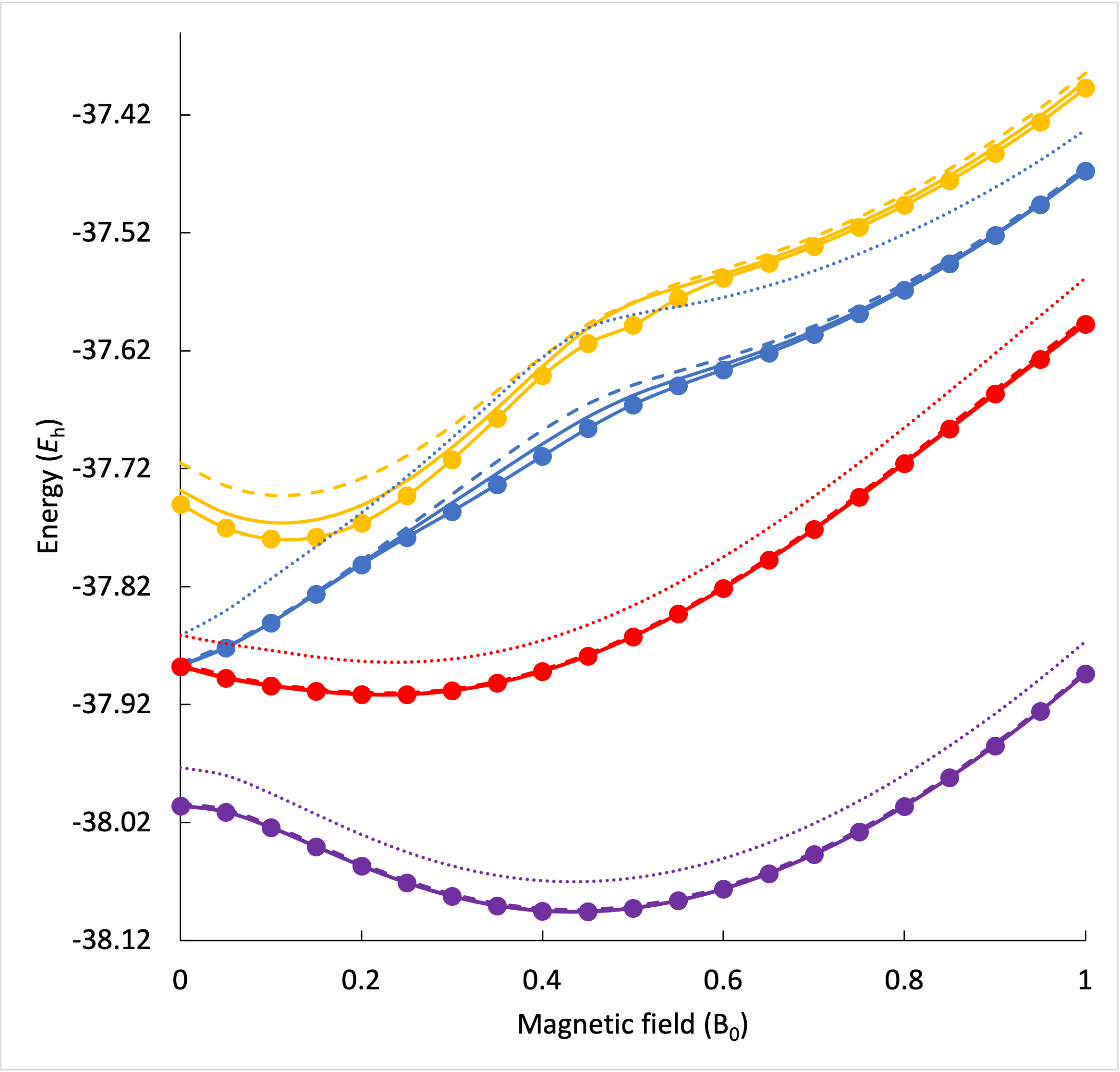}};
        \begin{scope}[x={(image.south east)},y={(image.north west)}]
            \node at (0.24,0.95) (angle) {\scalebox{0.9}{$\phi=60^\circ$}};
            \node at (0.55,0.22) (label1) {\scalebox{0.8}{$1{^1A} (^1\Sigma^+)$}};
            \node at (0.6,0.36) (label2) {\scalebox{0.8}{$2{^1A} (^1\Pi)$}};
            \node at (0.58,0.57) (label6) {\scalebox{0.8}{$3{^1A} (^1\Pi)$}};
            \node at (0.7,0.83) (label6) {\scalebox{0.8}{$4{^1A} (^1\Delta)$}};
        \end{scope}
    \end{tikzpicture}
    \end{minipage} \hspace{0.02\textwidth}
    \begin{minipage}[t]{0.47\textwidth}
    \begin{tikzpicture}
        \node[anchor=south west,inner sep=0] (image) at (0,0) {\includegraphics[width=\textwidth]{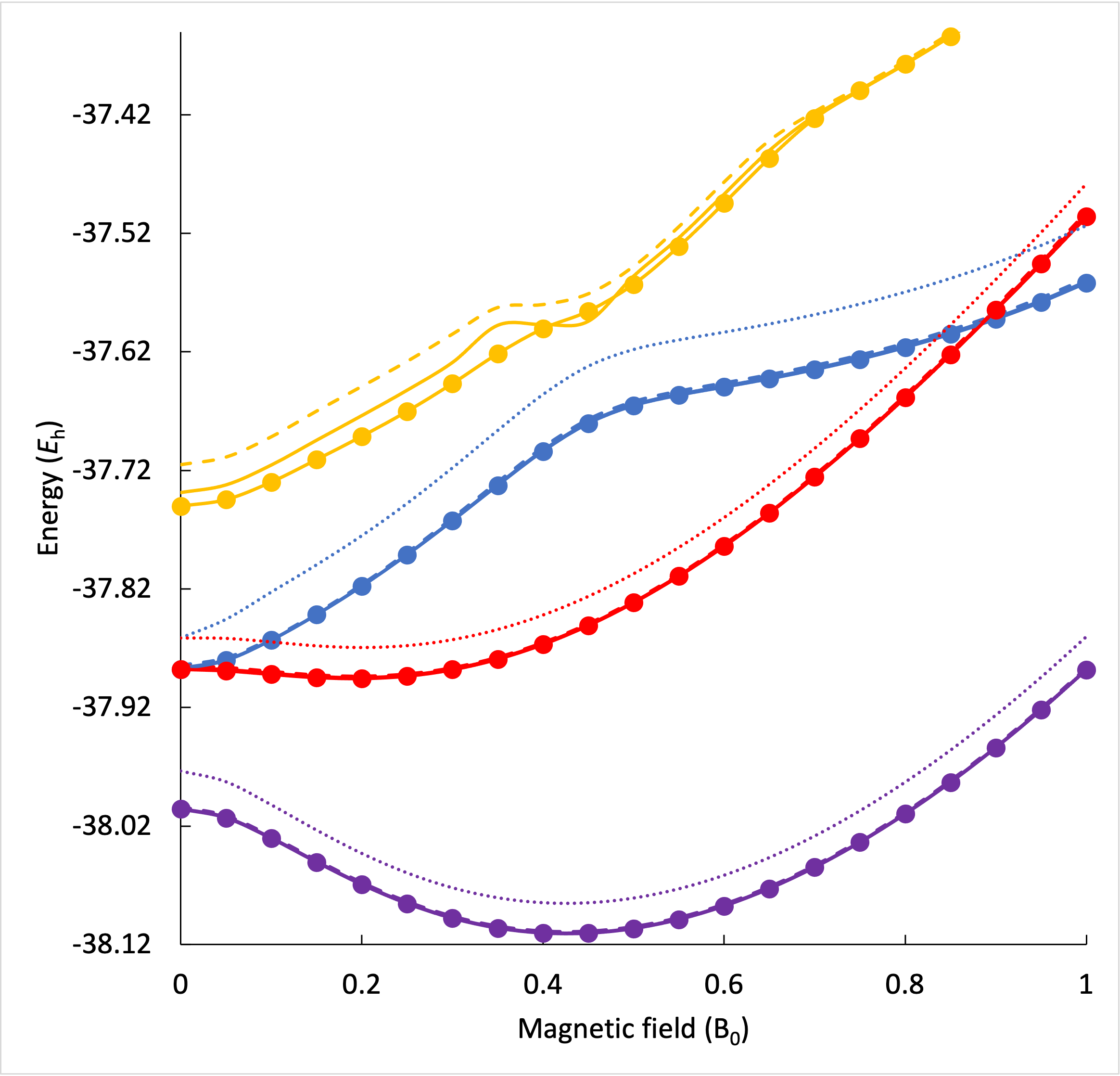}};
        \begin{scope}[x={(image.south east)},y={(image.north west)}]
            \node at (0.24,0.95) (angle) {\scalebox{0.9}{$\phi=90^\circ$}};
            \node at (0.55,0.2) (label1) {\scalebox{0.8}{$1{^1A}' (^1\Sigma^+)$}};
            \node at (0.6,0.38) (label2) {\scalebox{0.8}{$1{^1A}'' (^1\Pi)$}};
            \node at (0.56,0.55) (label6) {\scalebox{0.8}{$2{^1A}' (^1\Pi)$}};
            \node at (0.76,0.83) (label6) {\scalebox{0.8}{$3{^1A}' (^1\Delta)$}};
        \end{scope}
    \end{tikzpicture}
    \end{minipage}
\end{minipage}
\caption{The low-lying singlet states of CH$^+$ as a function of the magnetic-field strengths for different orientations of the molecule at the (EOM-)CC2 , CCSD, CC3 and CCSDT levels of theory with the  cc-pVDZ basis set. }
\label{fig:CH+}
\end{figure}

In ref.~\onlinecite{Hampe2020}, the CH$^+$ cation has been studied  at the CCSD and CCSDT levels of theory  for an increasing magnetic-field strength and various orientations. In that study, the CCSDT results were practically indistinguishable from the exact FCI predictions. Here, the approximate ff-CC2 and ff-CC3 methods are tested for the same system. The accuracy of the CC3 predictions is examined in comparison to the CCSDT level of theory, while the performance of the CC2 approximation is tested against CCSD.

Calculations were carried out at  the ground state field-free  CCSD/cc-pVDZ equilibrium distance of $2.1275\ \Bohr$  and are plotted in fig.~\ref{fig:CH+} against the magnetic-field strength. The $^1\Sigma^+$ ground state  (purple) which has a ${1\sigma}^2{2\sigma}^2{3\sigma}^2$ electronic configuration was chosen as reference for the subsequent EOM-CC calculations. The excited states that were targeted are the $^1\Pi$ state (red and blue) and the  $^1\Delta$ state (yellow). Both are doubly degenerate states which are described by a ${1\sigma}^2{2\sigma}^2{3\sigma}^1 1\pi^1$ and a ${1\sigma}^2{2\sigma}^2 1\pi^2$ configuration, respectively. The energetically higher-lying $^1\Delta$ state has a predominant double-excitation character with respect to the reference. As such it is not well described at the CCSD level, as already noted in ref.~\onlinecite{Hampe2020}. For an in-depth discussion on the behaviour of the molecular cation in the presence of the magnetic field see refs.~\onlinecite{Hampe2020} and \onlinecite{Kitsaras2023}. 

In general, CC2 is not able to describe states with predominant double-excitation character. For this reason, CC2 results for the $^1\Delta$ state (yellow) are absent in the calculations. Regarding states with predominant single-excitation character, CC2 energies exhibit a positive shift that reaches up to $3.6 \cdot 10^{-2} \ E_{\mathrm{h}}$ relative to CCSDT throughout the calculations. If the excitation does not acquire a double-excitation character, a rather reasonable description is obtained. The shift is more or less constant and does not influence the excitation energies nor the overall behaviour much.

In the upper left panel fig.~\ref{fig:CH+}, the results for the parallel magnetic-field orientation are presented.   Here, the reference and the predominantly singly-excited states are well behaved. For these states, the deviation of the CC2 results from CCSDT ($\sim 10^{-2} \ E_{\mathrm{h}}$) is about one order of magnitude larger than the CCSD results ($\sim 10^{-3} \ E_{\mathrm{h}}$). Energies at the CC3 level, however, have a mean deviation at the order of $\sim 10^{-4} \ E_{\mathrm{h}}$. For the $^1\Delta$ state, which is a state with a predominant double-excitation character, the mean error at the CC3 level drops from $3.6 \cdot 10^{-2} \ E_\mathrm{h}$ for CCSD by more than half to $1.2 \cdot 10^{-2} \ E_\mathrm{h}$. 

As noted in ref.~\onlinecite{Hampe2020},  the avoided crossings  that arise between the original $^1\Delta$ state and the other states in all non-parallel magnetic-field orientations result in a transfer of the double-excitation character. This is  problematic because a single state is not described with the same accuracy for  every  magnetic-field strength. In addition, the field strength at which the avoided crossing is encountered  is strongly method dependent.
Such a transfer is most clearly seen at an orientation of $30^\circ$, where between $0.1\textrm{ - }0.6\  B_0$, two avoided crossings occur between the $4{^1A}(^1\Delta)$ state (yellow) and the $3{^1A}(^1\Pi)$ state (blue).
Notably, CC2 proves to be particularly inappropriate for these cases as the curve for the state arising from $^1\Pi$ (blue) is  qualitatively very different compared to the predictions at the more accurate levels of theory. The reason for the poor description of this state at the CC2 level stems from the fact that the predominant double-excitation character cannot be described within CC2 and the avoided crossing is simply not found. Instead, the curve follows the higher-lying $4{^1A}$ state for field strengths greater than $0.1 \ B_0$. Hence, for CC2, the resulting artificial state (blue dotted curve) is a nonphysical combination of two different (physical) states. For CC2, this is a general problem for any avoided crossing involving states with single- and double- excitation character.

CC3 results on the other hand give a consistent description of the states in the non-parallel magnetic-field orientations and offer a significant improvement relative to the CCSD results. The results obtained around the avoided crossing at CC3 are much closer to the CCSDT predictions as compared to CCSD. Additionally, the deviation of the CC3 results relative to CCSDT is about half of the deviation of CCSD relative to CCSDT when a predominant double-excitation character is present. Specifically for the more complicated $3{^1A}$ and $4{^1A}$ cases (blue and yellow, respectively),  the mean deviation drops by one order of magnitude to $10^{-3} \ E_\mathrm{h}$ at CC3 compared to CCSD.  This observation is consistent with the results for the $^1\Delta$ state in the parallel orientation. In the case of a predominant single-excitation character, the mean deviation at the CC3 level relative to the CCSDT results is $5.7\cdot 10^{-4}\ E_\mathrm{h}$ with a maximum error of $1.6\cdot 10^{-3}\ E_\mathrm{h}$. Compared to CCSD, the mean deviation is one order of magnitude smaller and the maximum error is less than half. In fact, the CC3 results are practically indistinguishable from CCSDT when a single-excitation character is dominant.

The study of the CH$^+$ cation shows a problematic behavior of the CC2 method for systems where avoided crossings with states of relevant double-excitation character appear. In such cases, CC2 can yield non-physical results. Using CC3 as an approximate triples correction works rather well, even in cases where the state acquires a significant double-excitation character.

 \subsubsection{The CH radical}

\begin{figure}[b!]
\centering
\begin{minipage}[t]{0.48\textwidth}
    \begin{tikzpicture}
        \node[anchor=south west,inner sep=0] (image) at (0,0) {\includegraphics[width=\textwidth]{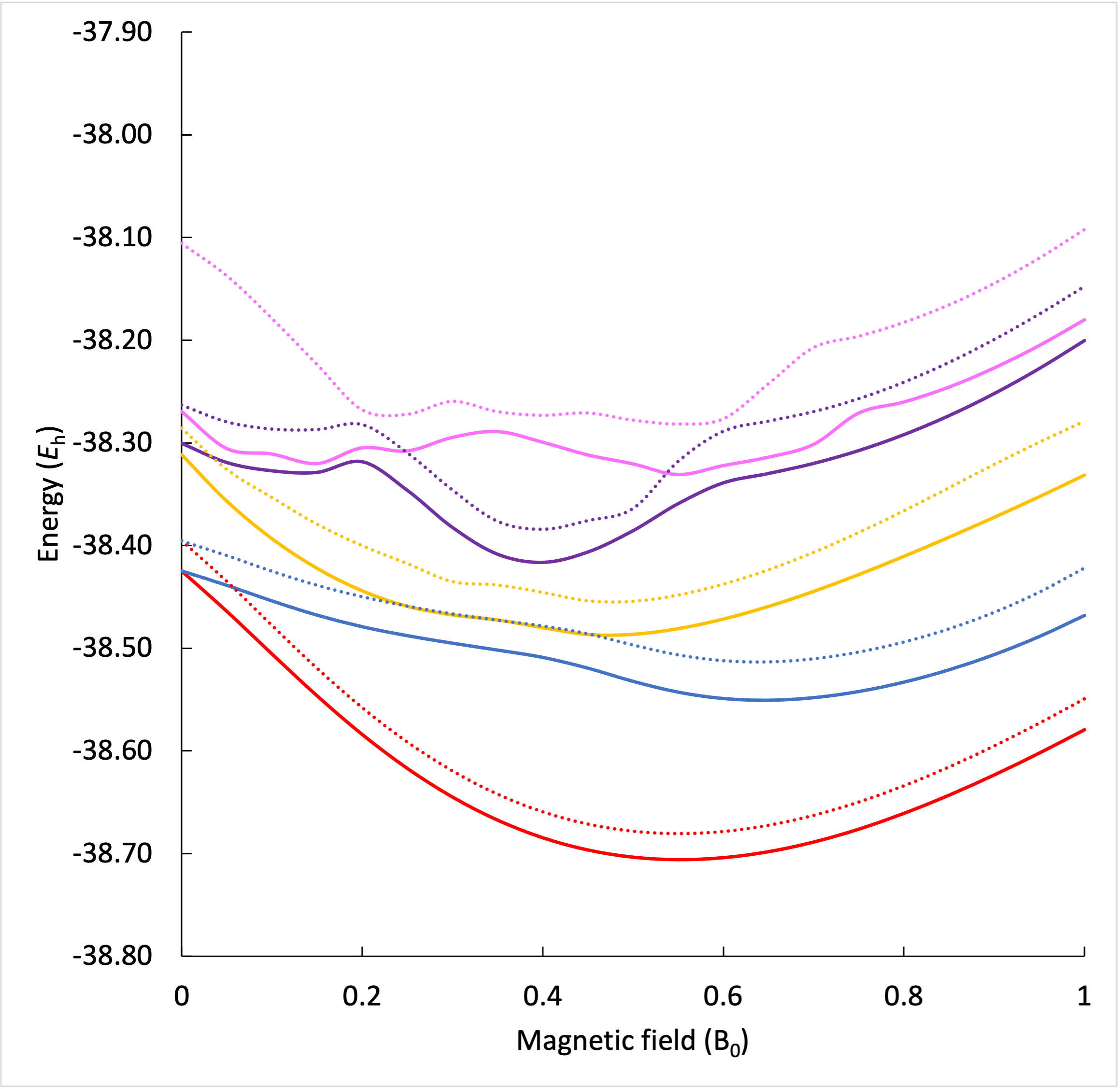}};
        \begin{scope}[x={(image.south east)},y={(image.north west)}]
            \node at (0.24,0.95) (angle) {\scalebox{0.9}{$\phi=60^\circ$}};
            \node at (0.5,0.95) (angle) {\scalebox{0.75}{CC2/CCSDT}};
            \node at (0.4,0.2) (label1) {\scalebox{0.8}{$1{^2A} (^2\Pi)$}};
            \node at (0.45,0.35) (label2) {\scalebox{0.8}{$2{^2A} (^2\Pi)$}};
            \node at (0.9,0.55) (label3) {\scalebox{0.8}{$3{^2A} (^2\Delta)$}};
            \node at (0.64,0.48) (label4) {\scalebox{0.8}{$4{^2A} (1{^2\Sigma}^-)$}};
            \draw[->] (label4)+(-0.00,0.02) -- +(-0.018,0.05);
            \node at (0.5,0.65) (label5) {\scalebox{0.8}{$5{^2A} (2{^2\Sigma}^-)$}};

            \node[anchor=west,minimum height = 0.5 cm] at (0.35,0.85)  {\scalebox{0.8}{CC2}};
            \draw[dotted] (0.35,0.85) -- +(-0.06,0) node [midway] (L) {};
            \node[anchor=west,minimum height = 0.5 cm] at (0.35,0.8)  {\scalebox{0.8}{CCSDT}};
            \draw[-] (0.35,0.8) -- +(-0.06,0) node [midway] (L) {};
        \end{scope}
    \end{tikzpicture}
    \end{minipage} \hspace{0.02\textwidth}
    \begin{minipage}[t]{0.48\textwidth}
    \begin{tikzpicture}
        \node[anchor=south west,inner sep=0] (image) at (0,0) {\includegraphics[width=\textwidth]{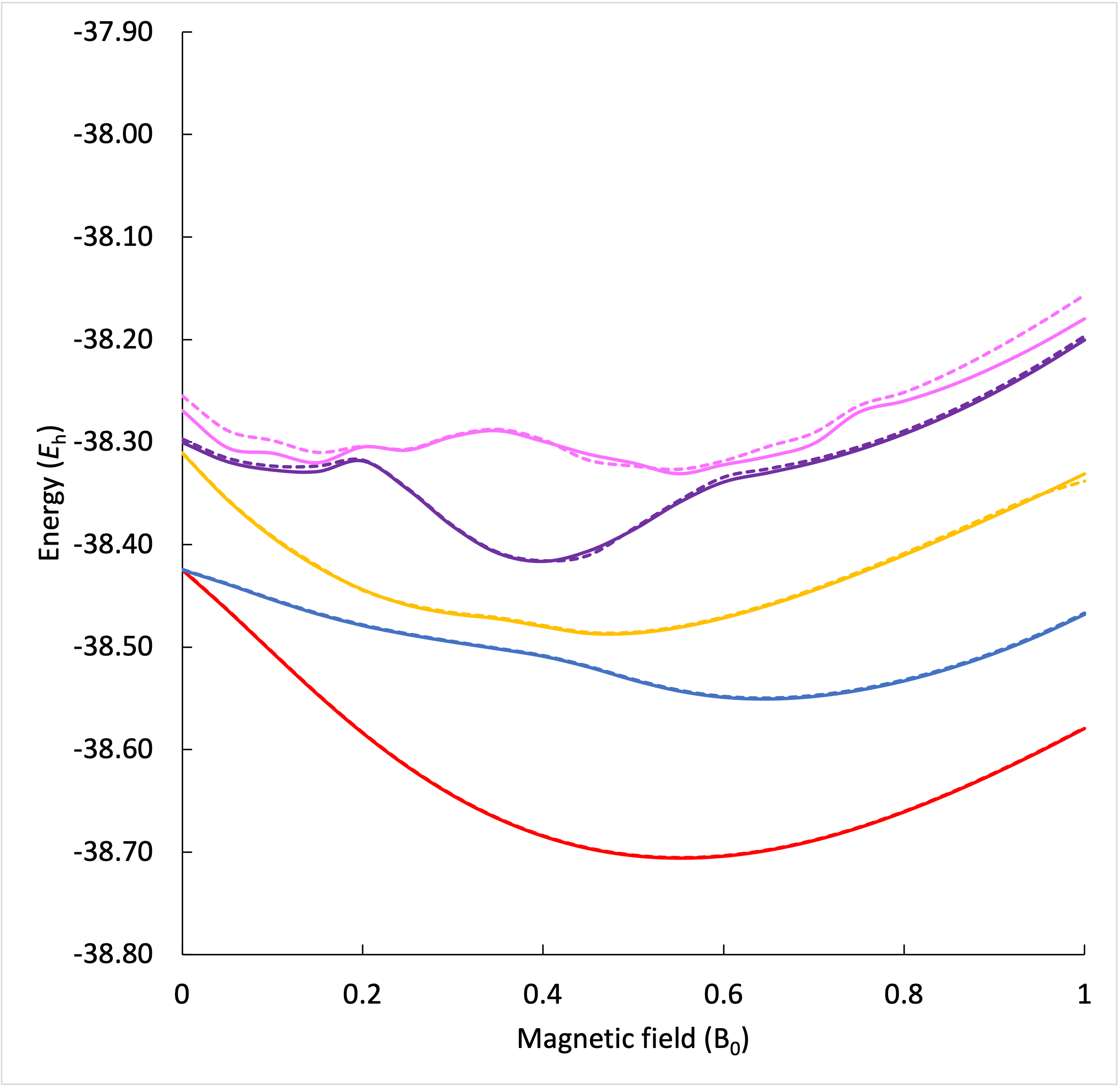}};
        \begin{scope}[x={(image.south east)},y={(image.north west)}]
            \node at (0.24,0.95) (angle) {\scalebox{0.9}{$\phi=60^\circ$}};
            \node at (0.5,0.95) (angle) {\scalebox{0.75}{CC3/CCSDT}};
            \node at (0.4,0.2) (label1) {\scalebox{0.8}{$1{^2A} (^2\Pi)$}};
            \node at (0.45,0.35) (label2) {\scalebox{0.8}{$2{^2A} (^2\Pi)$}};
            \node at (0.9,0.55) (label3) {\scalebox{0.8}{$3{^2A} (^2\Delta)$}};
            \node at (0.64,0.48) (label4) {\scalebox{0.8}{$4{^2A} (1{^2\Sigma}^-)$}};
            \draw[->] (label4)+(-0.00,0.02) -- +(-0.018,0.05);
            \node at (0.5,0.65) (label5) {\scalebox{0.8}{$5{^2A} (2{^2\Sigma}^-)$}};

            \node[anchor=west,minimum height = 0.5 cm] at (0.35,0.85)  {\scalebox{0.8}{CC3}};
            \draw[dashed] (0.35,0.85) -- +(-0.06,0) node [midway] (L) {};
            \node[anchor=west,minimum height = 0.5 cm] at (0.35,0.8)  {\scalebox{0.8}{CCSDT}};
            \draw[-] (0.35,0.8) -- +(-0.06,0) node [midway] (L) {};
        \end{scope}
    \end{tikzpicture}
    \end{minipage}

\caption{Comparison of CC2 (left panel) and CC3 (right panel) with CCSDT  results  obtained using the unc-cc-pVDZ basis set for four low-lying doublet states of the CH radical. The magnetic field is oriented at  a $60^\circ$ angle with respect to the molecular bond.   }
\label{fig:CH_60}
\end{figure}
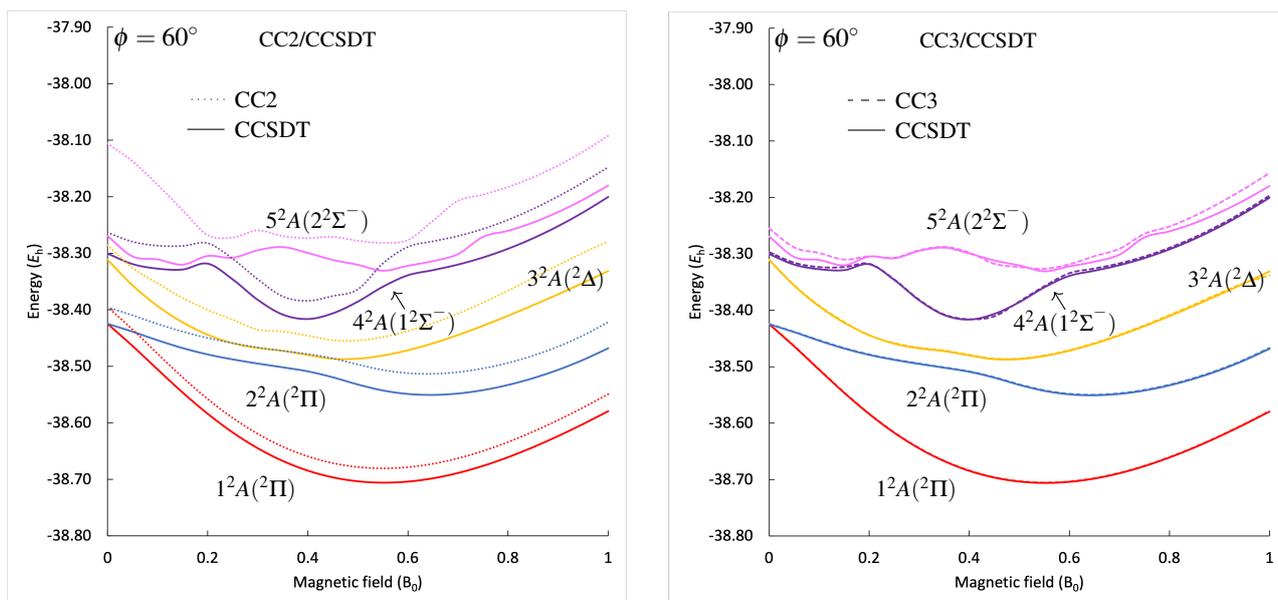

\begin{figure}[hp!]
\centering
\begin{minipage}{\textwidth}

    \begin{minipage}[t]{0.48\textwidth}
    \begin{tikzpicture}
        \node[anchor=south west,inner sep=0] (image) at (0,0) {\includegraphics[width=\textwidth]{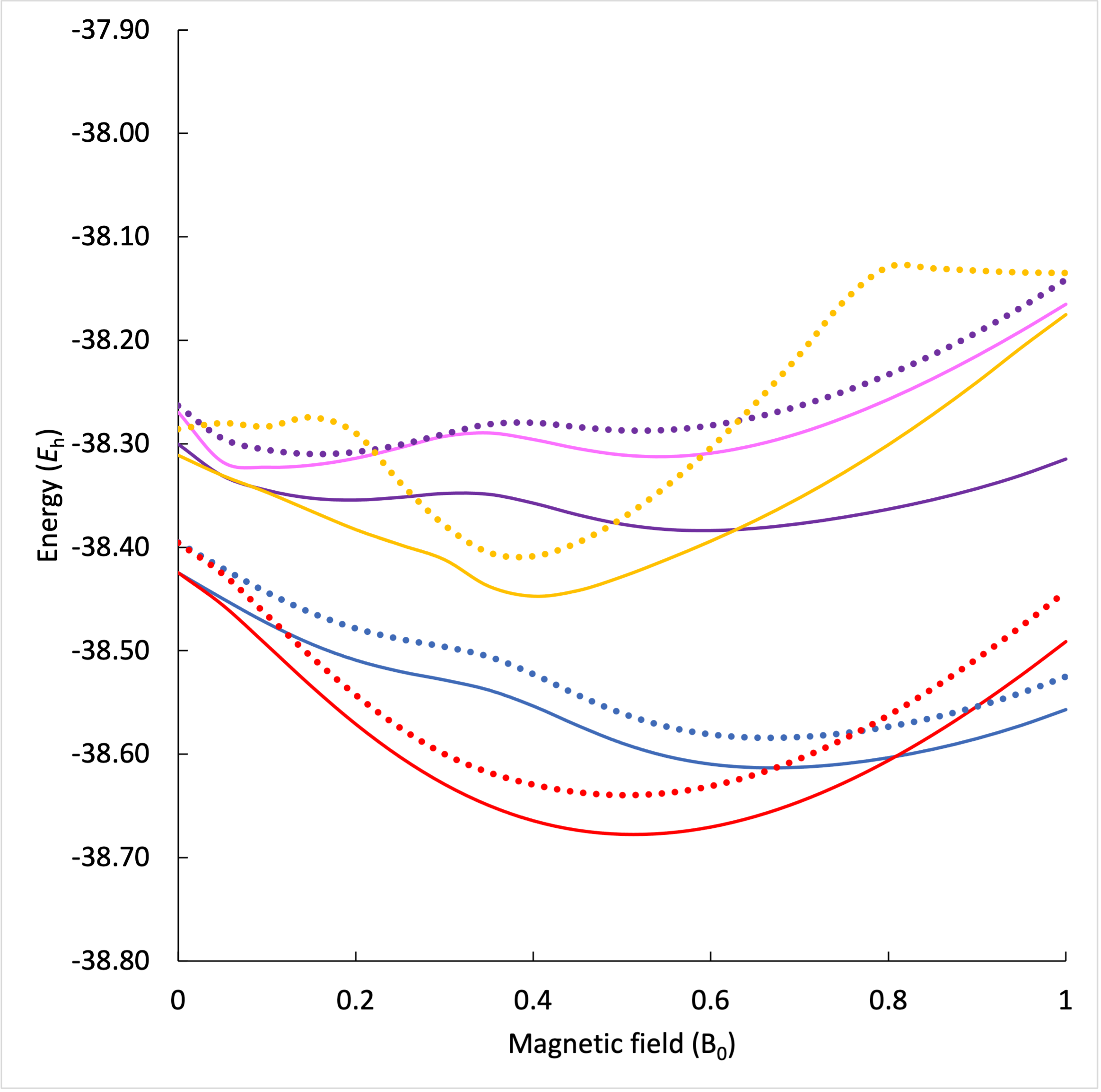}};
        \begin{scope}[x={(image.south east)},y={(image.north west)}]
            \node at (0.24,0.95) (angle) {\scalebox{0.9}{$\phi=90^\circ$}};
            \node at (0.85,0.95) (angle) {\scalebox{0.75}{$1{^2A}' (^2\Pi)$ reference }};
            \node at (0.5,0.95) (angle) {\scalebox{0.75}{CC2/CCSDT}};
            \node at (0.34,0.25) (label1) {\scalebox{0.8}{$1{^2A}'' (^2\Pi)$}};
            \node at (0.65,0.35) (label2) {\scalebox{0.8}{$1{^2A}' (^2\Pi)$}};
            \node at (0.64,0.45) (label3) {\scalebox{0.8}{$2{^2A}' (^2\Delta)$}};
            \node at (0.88,0.5) (label4) {\scalebox{0.8}{$2{^2A}'' (1{^2\Sigma}^-)$}};
            \node at (0.5,0.65) (label5) {\scalebox{0.8}{$3{^2A}'' (2{^2\Sigma}^-)$}};

            \node[anchor=west,minimum height = 0.5 cm] at (0.35,0.85)  {\scalebox{0.8}{CC2}};
            \draw[dotted] (0.35,0.85) -- +(-0.06,0) node [midway] (L) {};
            \node[anchor=west,minimum height = 0.5 cm] at (0.35,0.8)  {\scalebox{0.8}{CCSDT}};
            \draw[-] (0.35,0.8) -- +(-0.06,0) node [midway] (L) {};
        \end{scope}
    \end{tikzpicture}
    \end{minipage} \hspace{0.02\textwidth}
    \begin{minipage}[t]{0.48\textwidth}
    \begin{tikzpicture}
        \node[anchor=south west,inner sep=0] (image) at (0,0) {\includegraphics[width=\textwidth]{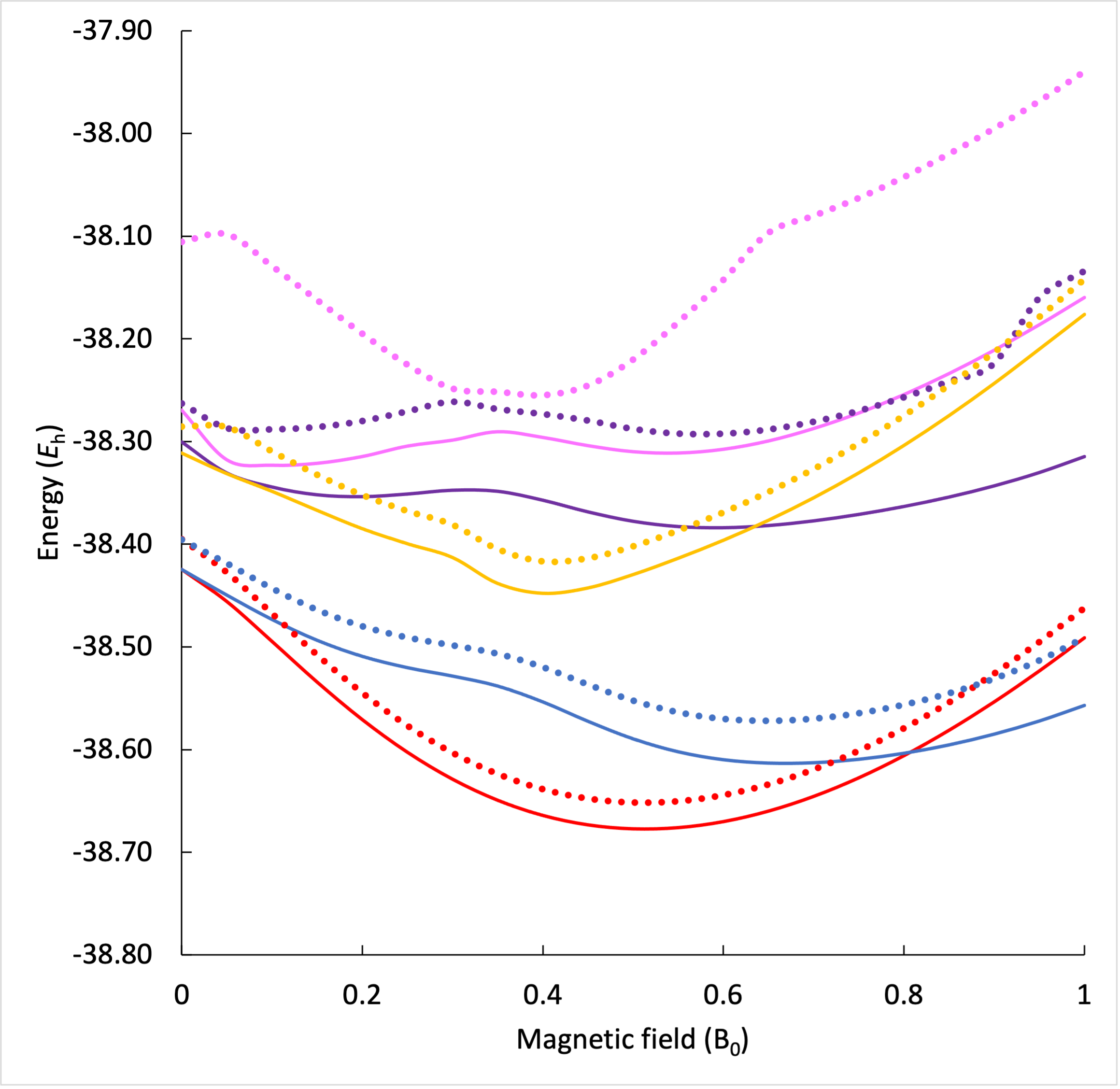}};
        \begin{scope}[x={(image.south east)},y={(image.north west)}]
            \node at (0.24,0.95) (angle) {\scalebox{0.9}{$\phi=90^\circ$}};
            \node at (0.85,0.95) (angle) {\scalebox{0.75}{$1{^2A}'' (^2\Pi)$ reference }};
            \node at (0.5,0.95) (angle) {\scalebox{0.75}{CC2/CCSDT}};
            \node at (0.34,0.25) (label1) {\scalebox{0.8}{$1{^2A}'' (^2\Pi)$}};
            \node at (0.65,0.35) (label2) {\scalebox{0.8}{$1{^2A}' (^2\Pi)$}};
            \node at (0.64,0.45) (label3) {\scalebox{0.8}{$2{^2A}' (^2\Delta)$}};
            \node at (0.88,0.5) (label4) {\scalebox{0.8}{$2{^2A}'' (1{^2\Sigma}^-)$}};
            \node at (0.5,0.65) (label5) {\scalebox{0.8}{$3{^2A}'' (2{^2\Sigma}^-)$}};
        \end{scope}
    \end{tikzpicture}
    \end{minipage}
\end{minipage}
\vspace{0.005\textwidth}
\centering

\begin{minipage}{\textwidth}

    \begin{minipage}[t]{0.48\textwidth}
    \begin{tikzpicture}
        \node[anchor=south west,inner sep=0] (image) at (0,0) {\includegraphics[width=\textwidth]{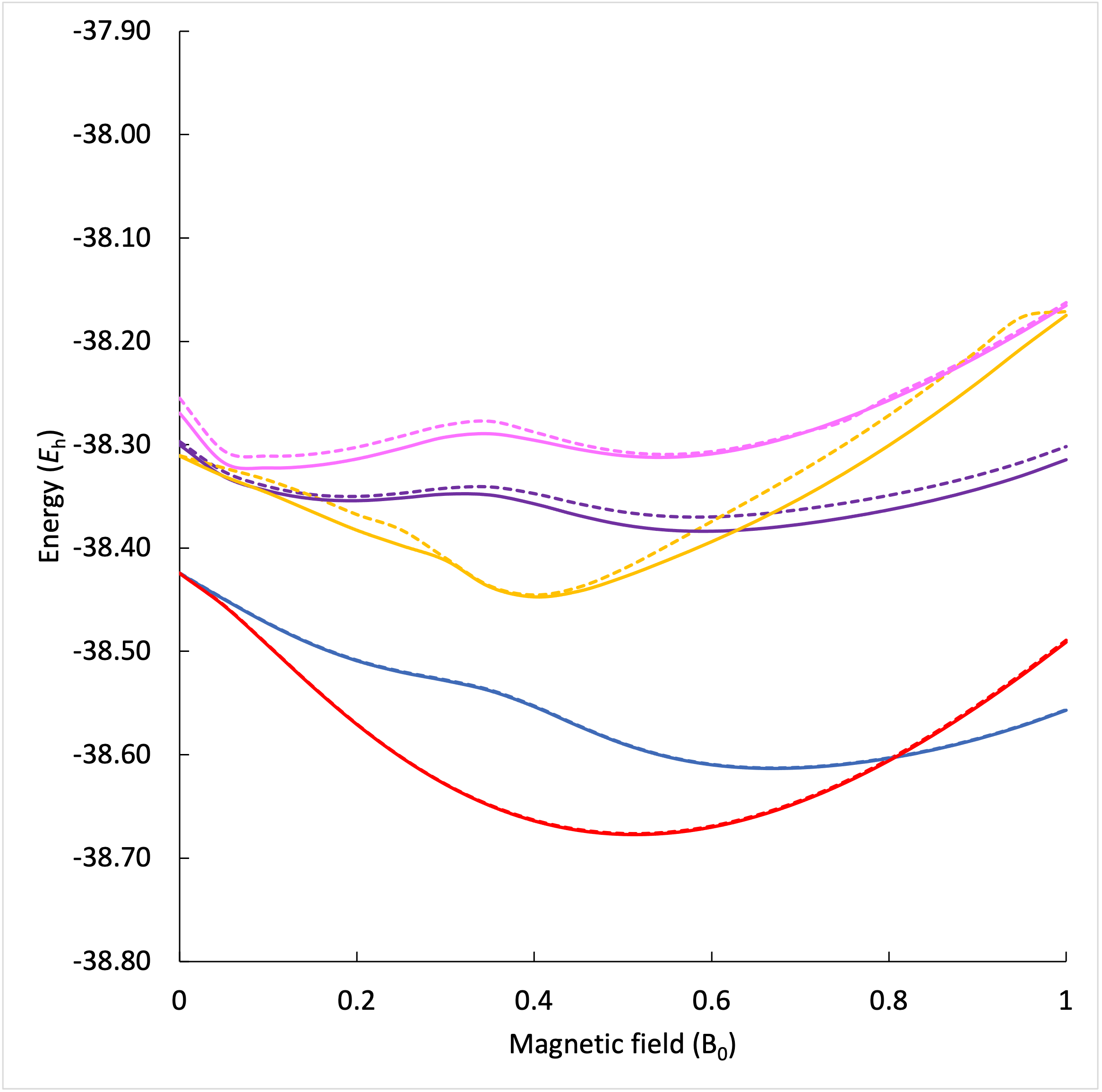}};
        \begin{scope}[x={(image.south east)},y={(image.north west)}]
            \node at (0.24,0.95) (angle) {\scalebox{0.9}{$\phi=90^\circ$}};
            \node at (0.85,0.95) (angle) {\scalebox{0.75}{$1{^2A}' (^2\Pi)$ reference }};
            \node at (0.5,0.95) (angle) {\scalebox{0.75}{CC3/CCSDT}};
            \node at (0.34,0.25) (label1) {\scalebox{0.8}{$1{^2A}'' (^2\Pi)$}};
            \node at (0.65,0.35) (label2) {\scalebox{0.8}{$1{^2A}' (^2\Pi)$}};
            \node at (0.64,0.45) (label3) {\scalebox{0.8}{$2{^2A}' (^2\Delta)$}};
            \node at (0.88,0.5) (label4) {\scalebox{0.8}{$2{^2A}'' (1{^2\Sigma}^-)$}};
            \node at (0.5,0.65) (label5) {\scalebox{0.8}{$3{^2A}'' (2{^2\Sigma}^-)$}};

            \node[anchor=west,minimum height = 0.5 cm] at (0.35,0.85)  {\scalebox{0.8}{CC3}};
            \draw[dashed] (0.35,0.85) -- +(-0.06,0) node [midway] (L) {};
            \node[anchor=west,minimum height = 0.5 cm] at (0.35,0.8)  {\scalebox{0.8}{CCSDT}};
            \draw[-] (0.35,0.8) -- +(-0.06,0) node [midway] (L) {};
        \end{scope}
    \end{tikzpicture}
    \end{minipage} \hspace{0.02\textwidth}
    \begin{minipage}[t]{0.48\textwidth}
    \begin{tikzpicture}
        \node[anchor=south west,inner sep=0] (image) at (0,0) {\includegraphics[width=\textwidth]{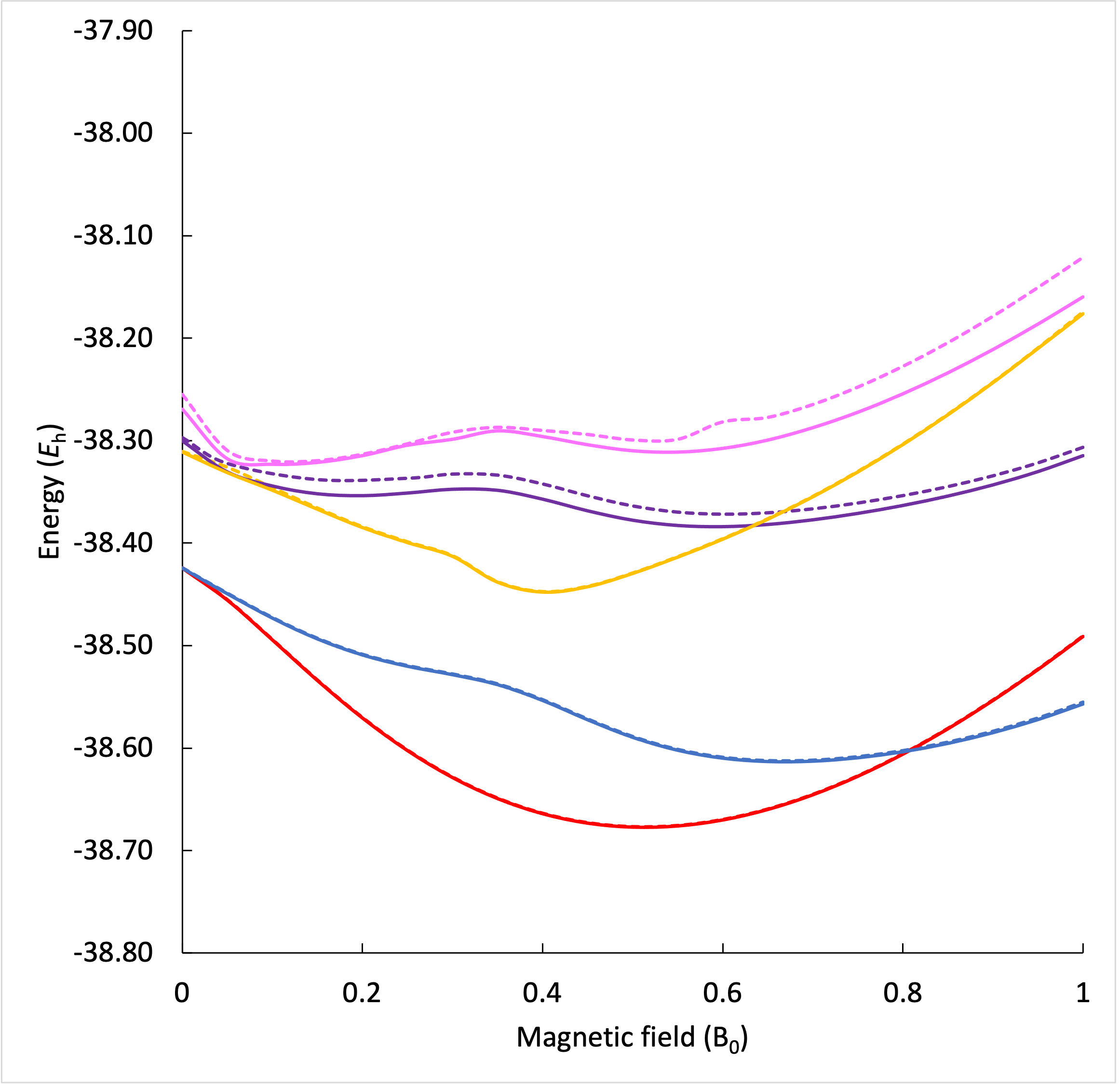}};
        \begin{scope}[x={(image.south east)},y={(image.north west)}]
            \node at (0.24,0.95) (angle) {\scalebox{0.9}{$\phi=90^\circ$}};
            \node at (0.85,0.95) (angle) {\scalebox{0.75}{$1{^2A}'' (^2\Pi)$ reference }};
            \node at (0.5,0.95) (angle) {\scalebox{0.75}{CC3/CCSDT}};
            \node at (0.34,0.25) (label1) {\scalebox{0.8}{$1{^2A}'' (^2\Pi)$}};
            \node at (0.65,0.35) (label2) {\scalebox{0.8}{$1{^2A}' (^2\Pi)$}};
            \node at (0.64,0.45) (label3) {\scalebox{0.8}{$2{^2A}' (^2\Delta)$}};
            \node at (0.88,0.5) (label4) {\scalebox{0.8}{$2{^2A}'' (1{^2\Sigma}^-)$}};
            \node at (0.5,0.65) (label5) {\scalebox{0.8}{$3{^2A}'' (2{^2\Sigma}^-)$}};
        \end{scope}
    \end{tikzpicture}
    \end{minipage}
\end{minipage}

\caption{Comparison of CC2 (upper panel) and CC3 (lower panel) with CCSDT  results   obtained using the  unc-cc-pVDZ basis set for the low-lying states of the CH radical with different reference states: $1{^2A}'$ (blue) left column, $1{^2A}''$ (red) right column. The magnetic-field orientation is perpendicular to the molecular bond.   }
\label{fig:CH_90}
\end{figure}

The CH radical was studied at  the \mbox{(EOM-)}CC2, CCSD, CC3 and CCSDT levels of theory using an unc-cc-pVDZ basis set at a field-free optimized bond length at the CCSD level of $2.1431\  \textrm{Bohr}$.  In fig.~\ref{fig:CH_60} and \ref{fig:CH_90}, a comparison of the CC2 and CC3 results with CCSDT for the skewed ($\phi=60^\circ$) and perpendicular ($\phi=90^\circ$) orientations, respectively, is presented. In these plots, the energy of the low-lying singlet states of the molecule is plotted as a function of the magnetic-field strength. 
Respective calculations for the parallel and $\phi=30^\circ$ orientations of the magnetic field relative to the molecular bond and results at the CCSD level can be found in ref.~\onlinecite{Kitsaras2023}.

Since the CH radical is an open-shell system with a doubly-degenerate $^2\Pi$ ground state  with a predominant ${1\sigma}^2{2\sigma}^2 {3\sigma}^2 1\pi^1$ electronic configuration, it constitutes a challenging test-case for the ff-EOM-CC treatment: Choosing one of the components of the $^2\Pi$ state as the reference state introduces an artificial break in their degeneracy, as only one of  the two $\Pi$ components is  treated as  an excited EOM-CC state. The severity of this effect is  quantified by the energy difference. It amounts to   about $\sim10^{-4} \ E_\mathrm{h}$ at the CC2 and CCSD levels of theory, about  $\sim10^{-5} \ E_\mathrm{h}$ for CC3, and about $\sim10^{-6} \ E_\mathrm{h}$ for CCSDT, showing that, as expected, the problem diminishes when improving the correlation treatment towards FCI.  The degenerate excited $^2\Delta$ state described by  a ${1\sigma}^2{2\sigma}^2 {3\sigma}^1 1\pi^2$ configuration was studied as well. For this state, one of the components (yellow) is characterised predominantly by a single excitation relative to the reference, while the other component is doubly excited and not targeted here.  Furthermore, two excited ${^2\Sigma}^-$ states were studied. The energetically lower-lying one $ 1 {^2\Sigma^-}$ has a predominant ${3\sigma}^\uparrow {1\pi^-}^\downarrow {1\pi^+}^\downarrow$ configuration (purple), while the second state $ 2 {^2\Sigma^-}$ (pink) has a multideterminental character with predominant $\left( {3\sigma}^\downarrow {1\pi^-}^\downarrow {1\pi^+}^\uparrow + {3\sigma}^\downarrow {1\pi^-}^\uparrow {1\pi^+}^\downarrow \right)$ configuration. Similar to the case of the $^2\Delta$ state, one of the determinants is doubly excited with respect to the reference. The description of the  $2 {^2\Sigma}^-$ state is particularly challenging. In principle, EOM approaches are  applicable to excited states with a multiconfigurational character. However, for such a description to work well, the strongly contributing determinants need to be singly excited with respect to the reference,\cite{Schreiber2008} which is  not the case here. Additionally, due to the symmetry breaking in the  CC reference state, the configurations no longer have the same weight. 

Overall, it is noted that  the performance of ff-CC2 is able to provide a qualitative description for states with  single-excitation character. It has large errors for excited states with mixed excitation character and completely fails to target the doubly-excited $^2\Delta$ component. Ff-CC3 on the other hand yields energies very close to CCSDT. 
Further discussion  in the following paragraphs on the performance of CC2 and CC3 is focused on the skewed $60^\circ$ and perpendicular $90^\circ$ magnetic-field orientations  relative to the molecular bond.

In fig.~\ref{fig:CH_60} (skewed $60^\circ$ orientation), the CC2 and CC3 energies can be  compared to the CCSDT reference. Results at the CC3 level exhibit  deviations from CCSDT at the order of  about $10^{-4}\ E_\mathrm{h}$ in the case of a predominant single-excitation character. These deviations are smaller by one order of magnitude as compared to CCSD. In the case of significant double-excitation character, the deviation from CCSDT is at the order of  $10^{-2} \ E_\mathrm{h}$. Compared to CCSD, the error is reduced by a factor of two. The CC2 method on the other hand still offers a crude qualitative description of the three lowest-energy states, but proves particularly problematic if not unphysical for the states arising from the $^2\Sigma^-$ states (purple and pink). The progression of the $4{^2A}$ (purple) and $5{^2A}$ (pink) states at the CC2 level  with increasing  magnetic-field strength is in many cases qualitatively very different  compared to the CCSDT reference and the more accurate methods, and apparently wrong. For example, the response of the $5{^2A}$ state (pink) to an increasing magnetic field at the CCSDT level is described as follows: The energy is relatively constant for field strengths between $B=0\textrm{ - }0.2 \ B_0$. It rises for stronger fields until $\sim 0.35\ B_0$ reaching a local maximum  due to an avoided crossing with a higher-lying state. The state  is then slightly stabilized up to $B=0.6 \ B_0$ after which the diamagnetic response becomes dominant. In addition, the state is  energetically close to the $4{^2A}$ state (purple). At the CC2 level, however, the $5{^2A}$ state (pink) starts off energetically higher in the field-free case with a very large error of $0.16 \ E_\mathrm{h}$ relative to CCSDT.  As seen in the left panel of fig.~\ref{fig:CH_60}, the development of the state is qualitatively different up to $B=0.2 \ B_0$: the state is gradually  stabilised  and approaches the CCSDT reference. For  $B \geq 0.2 \ B_0$, the CC2 deviation relative to CCSDT is lower by one order of magnitude ($\sim 10^{-2} \ E_\mathrm{h}$) compared to the field-free case.  Nonetheless, the energy development of the state exhibits features that are absent from the results from the more accurate methods, like two avoided crossings with the $4{^2A}$ (purple) state at $B=0.2 \ B_0$ and $B=0.6 \ B_0$. All theses qualitative differences constitute a strong indication that the character of the state at the CC2 level is at least over a range of field strengths qualitatively different compared to CCSDT and that the CC2 results are hence unphysical. The CC3 curve for this state on the other hand clearly follows the CCSDT reference over the complete range of field strengths. It can be noted that the deviations are small in the range between $0.2 \textrm{-} 0.6 \ B_0$, where a single-excitation character dominates.

The results for the  magnetic-field orientation perpendicular to the bond are shown in fig.~\ref{fig:CH_90}. Here, we compare two independent series of CC2 and CC3 calculations using different CC reference states, i.e., the two components of the field-free ground state $^2\Pi \rightarrow {^2A}',{^2A}''$ (blue and red, respectively) in order to judge the quality of the prediction. Again, the results are compared to  CCSDT reference calculations.  It is expected that the choice of reference should become less important for the EOM-CC states when approaching the FCI limit. This is indeed what is observed for the CCSDT results. They differ only slightly for the different choice of reference. In contrast, the CC2 method performs poorly for this test. Firstly,  it fails to consistently target all the states studied. For example, it was  not possible  to target the $3{^2A''}$ (pink) state using the $A'$ (blue) reference. Secondly, the results are in many cases qualitatively different when using different reference states.   Moreover, the magnetic-field strength at which the two components of the field-free ground state, i.e.,  the $1^2{A}'$ (blue) and $1^2{A}''$ (red) states, cross is not  predicted consistently at the CC2 level. The  states cross  at about $0.75\ B_0=180\ \mathrm{kT}$ when using $1^2{A}'$ (blue) as reference and at about $0.9\ B_0=210\ \mathrm{kT}$ when the  $1^2{A}''$ (red) is used. This amounts to a huge difference of $ 35\ \mathrm{kT}$. The CC3 approach on the other hand produces results very close to those at the CCSDT level. The two states arising from the $^2\Pi$ state are practically indistinguishable at the CC3 and CCSDT levels of theory. The difference at their crossing point when using the two references is about $0.05\ B_0=1\ \mathrm{kT}$ at CC3 and   smaller then $<0.01\ B_0=0.2\ \mathrm{kT}$ at CCSDT. The three higher-energy states are qualitatively in agreement between the two methods with larger deviations when a double-excitation character is present. On closer inspection, the  $3{^2A}''$ state (pink) is described slightly better with the $1^2{A}'$ (blue) reference, while the $2{^2A}'$ state (yellow) is described better with the $1^2{A}''$ (red) reference. This behaviour is due to contributions from doubly-excited determinants, that differ depending on the reference CC state. 

An analogous comparison for the parallel and $\phi=30^\circ$ orientation of the magnetic field is presented in ref.~\onlinecite{Kitsaras2023}. For the $30^\circ$ orientation, the observations are very similar to those at  $60^\circ$. For the parallel case, the performance of the CC$n$ approximations is more consistent among the different orientations as, due to symmetry, the mixing between the states is much more limited.

The calculations presented here consistently show that CC3 offers a satisfactory approximation to CCSDT even when a complex multiconfigurational character that arises from both single and double excitations is present. CC2  offers a crude qualitative description for the first few excited states that have a strong single-excitation character. Even for those cases, however, quantitative results, like the crossing point between states, are inconsistent and depend on which state has been chosen as the CC reference. For the states with a more complex multiconfigurational character, CC2 results are found to be unreliable.

\FloatBarrier

\subsection{Geometry optimizations in a strong magnetic field}
Geometry optimizations were performed on methane (CH$_4$) and ethylene (C$_2$H$_4$) in the presence of a magnetic field for different magnetic-field orientations and strengths. The unc-cc-pVTZ basis set was used to perform calculations at the ff-CCSD and ff-CC2 levels of theory.  A geometry optimization based on numerical gradients was carried out for a magnetic-field strength  up to $0.5 \ B_0$ using a  $0.05 \ B_0$ step. The optimizer module of CFOUR\cite{cfour,Matthews2020} was used, while the post-HF results were obtained using the QCUMBRE program\cite{QCUMBRE} in conjunction with an interface to CFOUR for the underlying ff-HF-SCF solution.

\subsubsection{Methane CH$_4$ \label{subsec:app_CH4}}

\begin{figure}[H]

     \centering
    \begin{minipage}[t]{0.4\textwidth}
        \centering
        \begin{tikzpicture}
          \draw  [->] (0.7,0.5) -- (0.7,1.5) node[anchor=east]{\footnotesize B};
          \node at(0,0){\chemfig{C([:90,1.7]-H)([:-19,2]-H)([:240,1.8]<H)([:200,1.5]<:H)}};
          \draw (-0.2,0.25) arc (90:-19:0.3) ;
          \node at(0.2,0.2) {\footnotesize $\omega$};
          \node at (-0.5,1) {\footnotesize $R_1$};
          \node at (0.6,-.7) {\footnotesize $R_2$};
        \end{tikzpicture}
        \caption{Methane with the magnetic field oriented parallel to one of the  C-H bonds. $C_3$ symmetry.}
        \label{fig:CH4_c3}
    \end{minipage}
    \hspace{0.02 \textwidth}
    \begin{minipage}[t]{0.4\textwidth}
        \centering
        \begin{tikzpicture}
          \draw  [->] (0,0.1) -- (0,1.5) ;
          \node at(0,1.5)  (B) [anchor=west] {\footnotesize B};
          \node at(0,-0.4){\chemfig{C([:40,2]-H)([:140,2]-H)([:-75,2]<H)([:-105,2]<:H)}};
          \draw (0.07,-0.5) arc (-76:-104:0.3) node[midway,sloped,below]{\footnotesize $\omega$};
          \node at (0.5,-1) {\footnotesize $R$};
        \end{tikzpicture}
        \caption{Methane with the magnetic field oriented parallel to the bisector of a H-C-H angle. $S_4$ symmetry.}
        \label{fig:CH4_s4}
    \end{minipage}
\end{figure}

 \begin{figure}[t!]
    $C_3$
    \centering{
    {%
    \setlength{\fboxsep}{0pt}%
    \setlength{\fboxrule}{0.5pt}%

    \fbox{\includegraphics[scale=0.36]{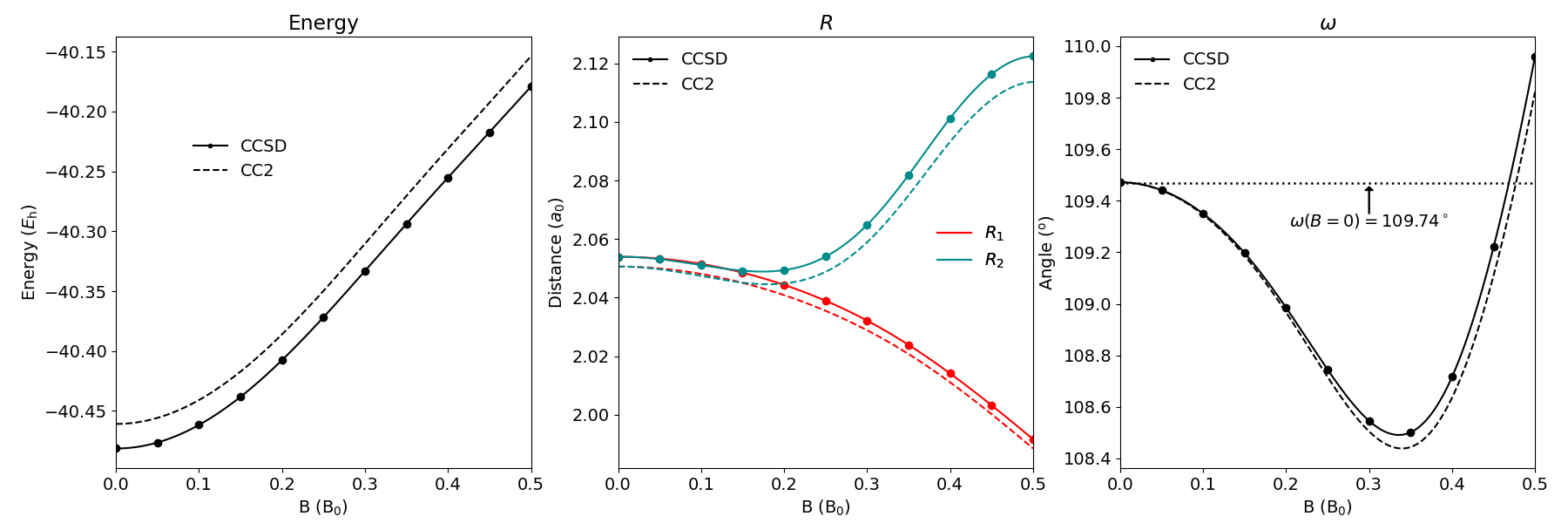}}%
    
    }%
    }
    
    \hfill
    
    $S_4$
    \centering{
    {%
    \setlength{\fboxsep}{0pt}%
    \setlength{\fboxrule}{0.5pt}%
    \fbox{\includegraphics[scale=0.36]{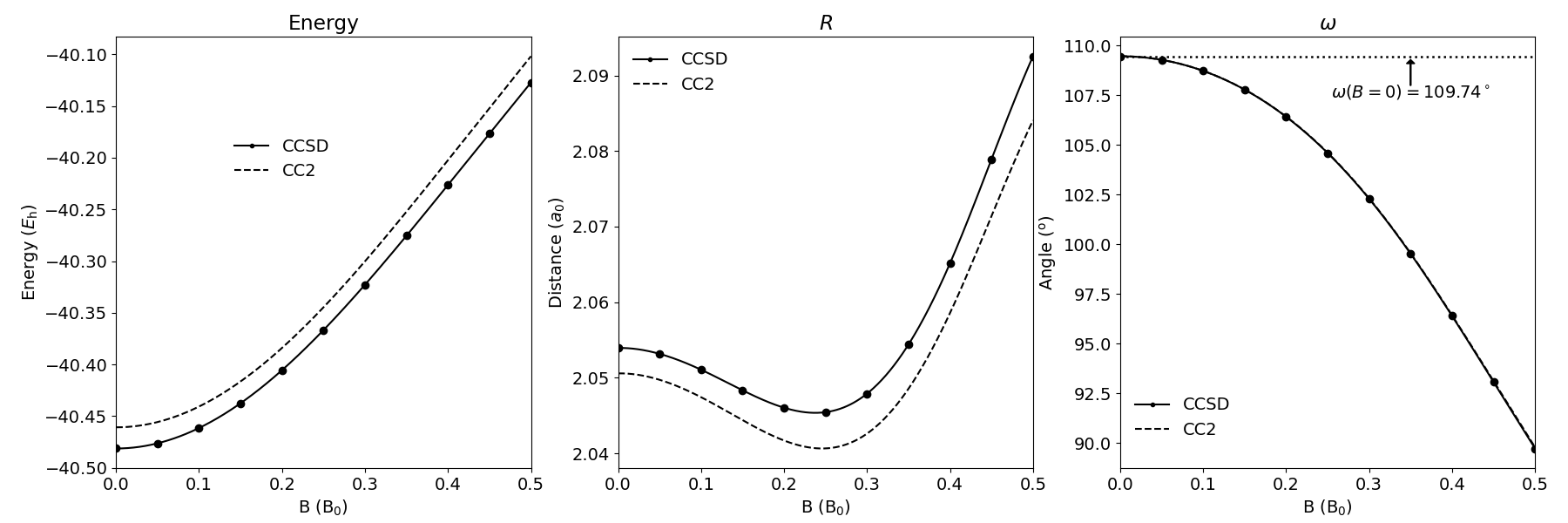}}%
    }%
    }
    \caption{Energies in $E_\mathrm{h}$ (left), as well as bond lengths in $\Bohr$ (middle) and angles in degrees (right) for the energetically lowest singlet state of the methane molecule at the optimized geometry at the CCSD (full line) and CC2 (dashed line) levels of theory as a function of the magnetic-field strength for two different orientations of the magnetic field with respect to the molecule: upper panel $B$ parallel to a C-H bond ($C_3$ symmetry), lower panel $B$ parallel to the bisector of the H-C-H angle ($S_4$ symmetry).}
    \label{fig:CH4_ground}

\end{figure}

In the field-free case, the symmetry of methane is described by the tetrahedral $T_d$ point group. Two orientations of the molecule relative to the magnetic field were studied. In the first one, one of the C-H bonds coincides with the magnetic field reducing the symmetry to $C_3$ as depicted in fig.~\ref{fig:CH4_c3}. In the second orientation, depicted in fig.~\ref{fig:CH4_s4}, the bisector of one of the H-C-H angles defines the magnetic-field orientation, which reduces the symmetry to $S_4$. In both cases, geometry optimizations were performed for the energetically lowest-lying singlet state $^1A(^1A_1)$.

The results for the geometry optmization, i.e., the energy at the constrained optimized geometry as well as the geometric parameters as a function of the magnetic-field strength are plotted in fig.~\ref{fig:CH4_ground}.  Regarding the $C_3$ case, the CC2 results for the geometry parameters reproduce the CCSD behaviour extremely well. At the CC2 level, bond lengths are about $0.004\ \Bohr$ shorter compared to CCSD predictions. The angles deviate by only $0.1^\circ$. In the $S_4$ case, similar trends are observed. The CC2 bond lengths are shorter by only $\sim 0.004\ \Bohr$ at CC2 compared to CCSD and angles deviate at most by $0.06^\circ$. The predictions for the geometrical parameters at the CC2 are thus nearly identical to those at CCSD. The energetical difference between the CCSD and CC2 results of around $22 \ mE_\mathrm{h}$ is also rather constant for different magnetic-field strengths for the two orientations.

The response of the molecular geometry to the increasing magnetic field together with a study of the excited states can be found in ref.\onlinecite{Kitsaras2023}.
Beyond these findings, methane has also been previously studied in ref.~\onlinecite{Pemberton2022} concerning  its stability and  the formation of exotic structures in the presence of a magnetic field at the ff-TDDFT level. In magnetic fields stronger than $0.5\ B_0$, a peculiar "fan-like" geometry  of CH$_4$ was predicted, resulting from a paramagnetic-bond formation in magnetic fields stronger than $0.5\ B_0$.\cite{Pemberton2022}

\FloatBarrier

\subsubsection{Ethylene CH$_2$=CH$_2$ \label{subsec:app_C2H4}}

\begin{figure}[H]
    \centering
    \begin{minipage}[t]{0.4\textwidth}
        \centering
        \tdplotsetmaincoords{100}{0}
        \begin{tikzpicture}[tdplot_main_coords]
            \centering
            \tdplotsetrotatedcoords{5}{0}{0}
            
            \draw[tdplot_rotated_coords,-] (-1,0,0.05) -- (1,0,0.05);
            \draw[tdplot_rotated_coords,-] (-1,0,-0.05) -- (1,0,-0.05);
            \draw[tdplot_rotated_coords,-] (-1,0,0) -- (-1.75,0,-1.299);
            \draw[tdplot_rotated_coords,-] (-1,0,0) -- (-1.75,0,1.299);
            \draw[tdplot_rotated_coords,-] (1,0,0) -- (1.75,0,1.299);
            \draw[tdplot_rotated_coords,-] (1,0,0) -- (1.75,0,-1.299);
    
            \foreach \z in {-1.5,-1,...,1.5}
		{
			\draw[tdplot_rotated_coords,draw=gray] (0,-3,\z) -- (0,3,\z);
			
		}
            \foreach \y in {-3,-2,...,3}
            {
                \draw[tdplot_rotated_coords,draw=gray] (0,\y,-1.5) -- (0,\y,1.5) ;
            }
    
            \node[tdplot_rotated_coords,fill=white] at (-1,0,0) (C1){C};
            \node[tdplot_rotated_coords,fill=white] at (1,0,0) (C2){C};
            \node[tdplot_rotated_coords,fill=white] at (-1.75,0,1.299) (H11){H};
            \node[tdplot_rotated_coords,fill=white] at (-1.75,0,-1.299) (H12){H};
            \node[tdplot_rotated_coords,fill=white] at (1.75,0,1.299) (H21){H};
            \node[tdplot_rotated_coords,fill=white] at (1.75,0,-1.299) (H22){H};
    
            \draw  [->] (-0.6,0,-0.7) -- (0.6,0,-0.7) node[anchor=north]{\footnotesize B};
            \tdplotsetrotatedcoords{-90}{90}{90}
            \tdplotdrawarc[tdplot_rotated_coords]{(C2)}{0.4}{160}{80}{}{};
            \node at(0.8,0,0.5) {\footnotesize $\omega$};
            \node at (-1.1,0,0.8) {\footnotesize $R_\textrm{H}$};
            \node at (-0.6,0,0.3) {\footnotesize $R_\textrm{C}$};
            \tdplotsetrotatedcoords{5}{0}{0}
            \node[tdplot_rotated_coords,text=gray] at (-0.2,0,1.9) {$\sigma_h$};
        \end{tikzpicture}

        \caption{Ethylene in a magnetic field oriented parallel to the C=C bond. $C_{2h}$ symmetry. The mirror plane $\sigma_h$ perpendicular to the magnetic field  is depicted by the gray grid.}
        \label{fig:C2H4_par}
    \end{minipage}
    \hspace{0.05\textwidth}
    \begin{minipage}[t]{0.4\textwidth}
        \centering
        \tdplotsetmaincoords{100}{0}
        \begin{tikzpicture}[tdplot_main_coords]
            \tdplotsetrotatedcoords{5}{0}{0}
    
            \draw[tdplot_rotated_coords,-] (-1,0,0.05) -- (1,0,0.05);
            \draw[tdplot_rotated_coords,-] (-1,0,-0.05) -- (1,0,-0.05);
            \draw[tdplot_rotated_coords,-] (-1,0,0) -- (-1.75,0,-1.299);
            \draw[tdplot_rotated_coords,-] (-1,0,0) -- (-1.75,0,1.299);
            \draw[tdplot_rotated_coords,-] (1,0,0) -- (1.75,0,1.299);
            \draw[tdplot_rotated_coords,-] (1,0,0) -- (1.75,0,-1.299);
            
            \node[tdplot_rotated_coords,fill=white,color=white] at (-1,0,0) (C1){C};
            \node[tdplot_rotated_coords,fill=white,color=white] at (1,0,0) (C2){C};
            \node[tdplot_rotated_coords,fill=white] at (-1.75,0,1.299) (H11){H};
            \node[tdplot_rotated_coords,fill=white] at (-1.75,0,-1.299) (H12){H};
            \node[tdplot_rotated_coords,fill=white] at (1.75,0,1.299) (H21){H};
            \node[tdplot_rotated_coords,fill=white] at (1.75,0,-1.299) (H22){H};

            \foreach \x in {-2.5,-2,...,2.5}
		{
			\draw[tdplot_rotated_coords,draw=gray] (\x,-2,0) -- (\x,2,0);
			
		}
            \foreach \y in {-2,-1.5,...,2}
            {
                \draw[tdplot_rotated_coords,draw=gray] (-2.5,\y,0) -- (2.5,\y,0);
            }
    
            \node[tdplot_rotated_coords] at (-1,0,0) (C1){C};
            \node[tdplot_rotated_coords] at (1,0,0) (C2){C};
    
            \draw  [tdplot_rotated_coords,->] (2,0,-0.6) -- (2,0,0.6) node[anchor=west]{\footnotesize B};
            \tdplotsetrotatedcoords{-90}{90}{90}
            \tdplotdrawarc[tdplot_rotated_coords]{(C2)}{0.4}{160}{80}{}{};
            \node at(0.8,0,0.5) {\footnotesize $\omega$};
            \node at (-1.1,0,0.8) {\footnotesize $R_\textrm{H}$};
            \node at (-0.,0,0.5) {\footnotesize $R_\textrm{C}$};
            \tdplotsetrotatedcoords{5}{0}{0}
            \node[tdplot_rotated_coords,color=gray] at (-2.85,0,0) {$\sigma_h$};
          
        \end{tikzpicture}

        \caption{Ethylene in a magnetic field parallel to the  molecular plane and oriented perpendicular to the C=C bond. $C_{2h}$ symmetry. The mirror plane $\sigma_h$ perpendicular to the magnetic field  is depicted by the gray grid.}
        \label{fig:C2H4_inpl}
    \end{minipage}

\end{figure}

\begin{figure}[!t]

    parallel
    \centering{
    {%
    \setlength{\fboxsep}{0pt}%
    \setlength{\fboxrule}{0.5pt}%
    \fbox{\includegraphics[scale=0.36]{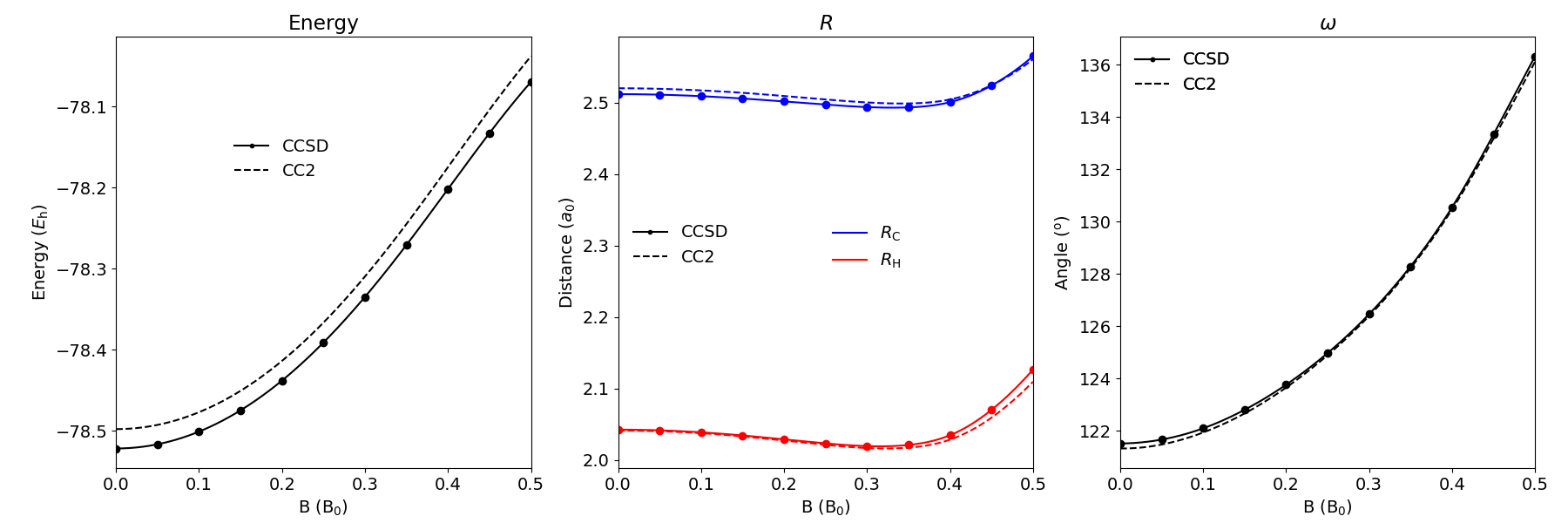}}%
    }%
    }
    
    \hfill
    
    perpendicular
    \centering{
    {%
    \setlength{\fboxsep}{0pt}%
    \setlength{\fboxrule}{0.5pt}%
    \fbox{\includegraphics[scale=0.36]{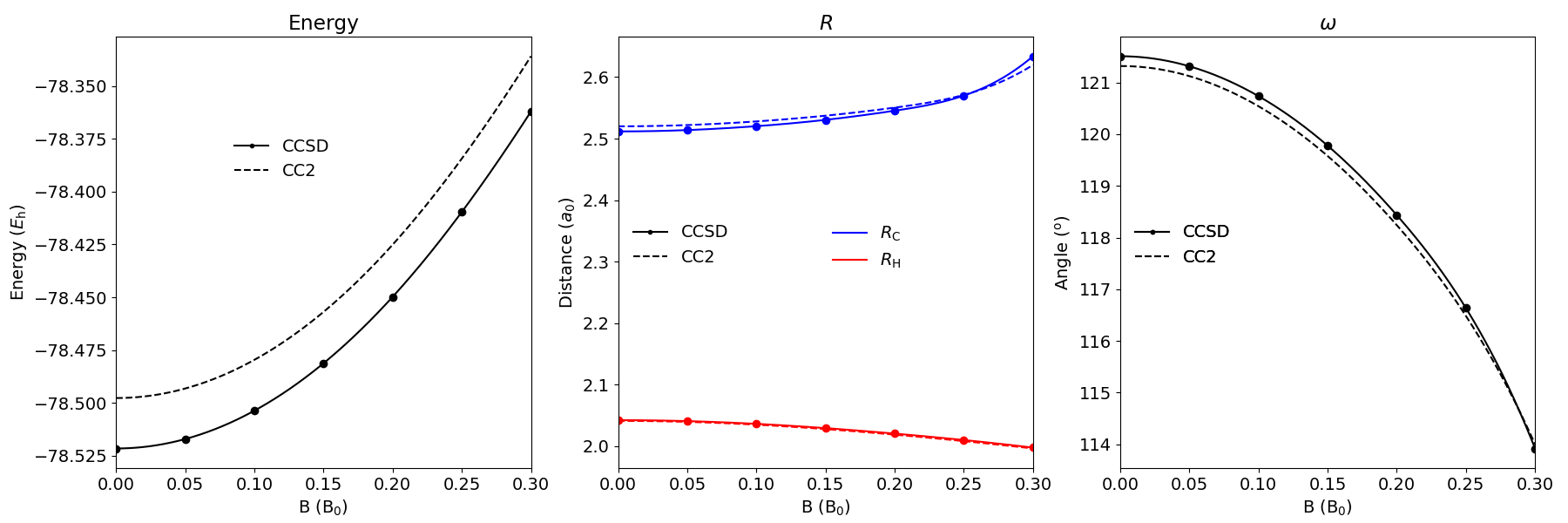}}%
    }%
    }
    \caption{Energies in $E_\mathrm{h}$ (left), as well as bond lengths in $\Bohr$ (middle) and angles in degrees (right) for the energetically lowest singlet state of the ethylene molecule at the optimized geometry at the CCSD (full line) and CC2 (dashed line) levels of theory as a function of the magnetic-field strength for two different orientations of the magnetic field with respect to the molecule: upper panel $B$ parallel to the C=C bond, lower panel $B$ in-plane perpendicular to the C=C bond.}
    \label{fig:C2H4_ground}

\end{figure}

Two magnetic-field orientations were studied in the case of ethylene, which are depicted in figs.~\ref{fig:C2H4_par} and \ref{fig:C2H4_inpl}. In both cases the symmetry is reduced from $D_{2h}$ in the absence of a magnetic field to two distinct  $C_{2h}$ subgroups. In the first case, the magnetic field is oriented parallel to the C=C bond and in the second case it is  perpendicular to the C=C bond and lies within the molecular plane. The gray grid in the figures depicts the different mirror planes. In spite of the symmetry reduction, the hydrogen and carbon nuclear centers remain symmetry equivalent. In regard to the constrained geometry optimization for the energetically lowest-lying singlet state $^1A_g(^1A_g)$, the degrees of freedom are the C=C and C-H bond lengths, $R_{\mathrm{C}}$ and $R_{\mathrm{H}}$, respectively, together with the C-C-H angle $\omega$.

The energy at the optimized geometry and the geometric parameters of the molecule are plotted for the two different orientations as a function of the magnetic field in fig.~\ref{fig:C2H4_ground}. The CC2 and CCSD predictions are in qualitative agreement to each other. As expected, the CC2 total energies are higher than the CCSD energies for different magnetic-field strengths and orientations by a nearly constant shift of $0.025\ E_\mathrm{h}$. Similarly to the methane study, the geometric parameters only show minimal deviation between the two approaches, with a mean deviation of  $0.006\ \textrm{Bohr}$ for the bond lengths and  $0.1 ^\circ$ for the angle. Noteworthy is the dissociation of the molecule for field strengths greater than $0.30\ B_0$ in the perpendicular orientation of the magnetic field. This prediction is consistent both at the CC2 and CCSD levels of theory.

A detailed discussion of the response of the geometry to the magnetic field together with an investigation of the spectrum in the presence of a magnetic field for ethylene can be found in ref.~\onlinecite{Kitsaras2023}.

The calculations for methane and ethylene show that the CC2 method performs very well against CCSD for geometry optmizations. The geometrical parameters obtained at the CC2 level are practically indistinguishable from the CCSD results. 

\FloatBarrier
\section{Conclusion \label{sec:concl}}
 In this paper, the CC2 and CC3 methods, as well as the respective EOM-CC approach have been implemented for calculations of ground and excited states of  molecules in finite magnetic fields. These methods approximate the standard CC truncations, i.e., CCSD and CCSDT, respectively, and lower the scaling by one order of magnitude relative to their parent methods.

Various systems were studied in the presence of a magnetic field using these approximate CC$n$ methods and were compared to calculations using the standard CC truncations to test their performance. The triplet states of the Mg atom were studied at the CCSD, CC3 and CCSDT levels of theory. B-$\lambda$ curves were generated using an adaptation of a previously reported extrapolation scheme.\cite{Hampe2020} These calculations have been essential to the assignment of a spectrum from a strongly magnetic White Dwarf star.\cite{Hollands2023} The CC3 approach enabled calculations with larger basis sets, that give an accuracy similar to CCSDT. 
Additionally, the diatomic cation CH$^+$ and radical CH, which are candidate molecules to occur in the atmosphere of strongly magnetic White Dwarfs, were studied at the (EOM-)CC2, (EOM-)CCSD, (EOM-)CC3 and (EOM-)CCSDT levels of theory in various magnetic-field strengths and orientations. The CC2 approach was found to yield non-physical results when avoided crossings between excited states with a predominant double-excitation character are involved. CC3 on the other hand is shown to  replicate the CCSDT behavior for ground and excited states with a predominant single-excitation character. It  even achieves  to capture a large extent of the double-excitation character.
Geometry optimizations were performed for the energetically lowest-lying singlet state of the small organic molecules CH$_4$ and C$_2$H$_4$  in different highly symmetric magnetic-field orientations and different field strengths at the CC2 and CCSD levels of theory. The study of these systems contributes to a better understanding of the geometry response of small molecules to the magnetic field. Optimized geometry parameters at the CC2 level of theory practically replicate the CCSD results. 

The results in this paper show that CC2 may be a good alternative to CCSD for larger systems, particularly for geometry optmizations. On the other hand, EOM-CC2 suffers from the inability to account for excited states with a double-excitation character. This can lead to unphysical results, when avoided crossings with such states are involved. The latter, however, is more the rule than the exception for species in strong magnetic fields, particularly, but not only, in skewed orientations of the field. For this reason, EOM-CC2 is only applicable if the presence of avoided crossings of this kind are ensured not to contribute.

The CC3 method proves to have merits in finite-magnetic field calculations. For both ground and  excited states with predominant single-excitation character, results obtained at the (EOM-)CC3 level of theory are essentially indistinguishable from the (EOM-)CCSDT results. In addition, EOM-CC3 offers a significant improvement relative to EOM-CCSD for excited states with predominant double-excitation character with results closer qualitatively to full EOM-CCSDT. The more favourable scaling of $N^7$ as compared to $N^8$ of CCSDT allows an approximate treatment of triple corrections and high-accuracy predictions for larger systems with larger basis sets. 

\section{Acknowledgments}
The authors thank  Prof. Dr. Jürgen Gauss for valuable discussions. This work has been supported by the Deutsche Forschungsgmeinschaft under Grant STO 1239/1-1. 

\section{Data Availability}
The data that support the findings of this study are available within the article and its supplementary material.

\printbibliography

@book{M.R.Manaa2005,
doi = {10.1016/B978-0-444-51766-1.X5000-8},
editor = {{M.R. Manaa}},
isbn = {9780444517661},
publisher = {Elsevier},
title = {{Chemistry at Extreme Conditions}},
url = {https://linkinghub.elsevier.com/retrieve/pii/B9780444517661X50008},
year = {2005}
}

@book{Fridman2008,
address = {Cambridge},
author = {Fridman, Alexander},
doi = {10.1017/CBO9780511546075},
isbn = {9780511546075},
publisher = {Cambridge University Press},
title = {{Plasma Chemistry}},
url = {https://www.cambridge.org/core/product/identifier/9780511546075/type/book},
year = {2008}
}

@article{Miao2020,
author = {Miao, Maosheng and Sun, Yuanhui and Zurek, Eva and Lin, Haiqing},
doi = {10.1038/s41570-020-0213-0},
issn = {2397-3358},
journal = {Nat. Rev. Chem.},
month = {oct},
number = {10},
pages = {508--527},
title = {{Chemistry under high pressure}},
url = {https://www.nature.com/articles/s41570-020-0213-0},
volume = {4},
year = {2020}
}

@book{Motzfeldt2012,
address = {Chichester, UK},
author = {Motzfeldt, Ketil},
doi = {10.1002/9781118457795},
isbn = {9781118457795},
month = {dec},
publisher = {John Wiley \& Sons, Ltd},
title = {{High Temperature Experiments in Chemistry and Materials Science}},
url = {http://doi.wiley.com/10.1002/9781118457795},
year = {2012}
}

@article{Schmelcher2012,
author = {Schmelcher, Peter},
doi = {10.1126/science.1224869},
issn = {0036-8075},
journal = {Science (80).},
month = {jul},
number = {6092},
pages = {302--303},
title = {{Molecule Formation in Ultrahigh Magnetic Fields}},
url = {https://www.sciencemag.org/lookup/doi/10.1126/science.1224869 https://www.science.org/doi/10.1126/science.1224869},
volume = {337},
year = {2012}
}

@article{Lange2012,
author = {Lange, K. K. and Tellgren, E. I. and Hoffmann, M. R. and Helgaker, T.},
doi = {10.1126/science.1219703},
issn = {0036-8075},
journal = {Science},
month = {jul},
number = {6092},
pages = {327--331},
title = {{A Paramagnetic Bonding Mechanism for Diatomics in Strong Magnetic Fields}},
url = {https://www.sciencemag.org/lookup/doi/10.1126/science.1219703},
volume = {337},
year = {2012}
}

@article{Schmelcher1997,
author = {Schmelcher, P and Cederbaum, L S and {P. Schmelcher} and {L. S. Cederbaum}},
journal = {Int. J. Quantum Chem.},
number = {5},
pages = {501--511},
publisher = {John Wiley {\&} Sons, Inc},
title = {{Molecules in strong magnetic fields: Some perspectives and general aspects}},
url = {https://onlinelibrary.wiley.com/terms-and-conditions},
volume = {64},
year = {1997}
}

@article{Tellgren2009,
author = {Tellgren, Erik I. and Helgaker, Trygve and Soncini, Alessandro},
doi = {10.1039/b822262b},
issn = {1463-9076},
journal = {Phys. Chem. Chem. Phys.},
number = {26},
pages = {5489},
title = {{Non-perturbative magnetic phenomena in closed-shell paramagnetic molecules}},
url = {http://xlink.rsc.org/?DOI=b822262b},
volume = {11},
year = {2009}
}

@article{Margrave2022,
author = {Margrave, John L.},
%doi = {10.1036/1097-8542.318500},
journal = {AccessScience},
month = {jan},
publisher = {McGraw-Hill Education},
title = {{High-temperature chemistry}},
url = {https://www.accessscience.com/content/article/a318500},
year = {2020}
}

@article{Schmelcher1991,
author = {Schmelcher, P. and Cederbaum, L. S.},
doi = {10.1002/QUA.560400836},
issn = {1097461X},
journal = {Int. J. Quantum Chem.},
number = {25 S},
pages = {371--385},
title = {{On molecules and ions in strong magnetic fields}},
volume = {40},
year = {1991}
}

@article{Rosner1984,
author = {Rosner, W. and Wunner, G. and Herold, H. and Ruder, H.},
doi = {10.1088/0022-3700/17/1/010},
issn = {0022-3700},
journal = {J. Phys. B At. Mol. Phys.},
month = jan,
number = {1},
pages = {29--52},
publisher = {IOP Publishing},
title = {{Hydrogen atoms in arbitrary magnetic fields. I. Energy levels and wavefunctions}},
url = {https://iopscience.iop.org/article/10.1088/0022-3700/17/1/010 https://iopscience.iop.org/article/10.1088/0022-3700/17/1/010/meta},
volume = {17},
year = {1984}
}

@article{Schmelcher1988,
    title = {{Electronic and nuclear motion and their couplings in the presence of a magnetic field}},
    year = {1988},
    journal = {Phys. Rev. A},
    author = {Schmelcher, P. and Cederbaum, L. S. and Meyer, H. D.},
    number = {12},
    pages = {6066--6079},
    volume = {38},
    doi = {10.1103/PhysRevA.38.6066},
    issn = {10502947}
}

@article{Schmelcher1988a,
author = {Schmelcher, P. and Cederbaum, L. S. and Meyer, H. D.},
doi = {10.1088/0953-4075/21/15/005},
issn = {0953-4075},
journal = {J. Phys. B At. Mol. Opt. Phys.},
month = {aug},
number = {15},
pages = {L445--L450},
title = {{On the validity of the Born-Oppenheimer approximation in magnetic fields}},
url = {https://iopscience.iop.org/article/10.1088/0953-4075/21/15/005},
volume = {21},
year = {1988}
}

@article{Schmelcher1988b,
author = {Schmelcher, P. and Cederbaum, L. S.},
doi = {10.1103/PhysRevA.37.672},
issn = {0556-2791},
journal = {Phys. Rev. A},
month = feb,
number = {3},
pages = {672--681},
title = {{Molecules in strong magnetic fields: Properties of atomic orbitals}},
url = {https://link.aps.org/doi/10.1103/PhysRevA.37.672},
volume = {37},
year = {1988}
}

@ARTICLE{Jordan1998,
    author = {{Jordan}, S. and {Schmelcher}, P. and {Becken}, W. and {Schweizer}, W.  },
    title = {Evidence for helium in the magnetic white dwarf GD 229},
  journal = {Astron. Astrophys.},
    year = {1998},
    month = aug,
    volume = 336,
    pages = {L33--L36},
}

@article{Soncini2004,
author = {Soncini, A. and Fowler, P.W.},
doi = {10.1016/j.cplett.2004.10.110},
issn = {00092614},
journal = {Chem. Phys. Lett.},
month = dec,
number = {1-3},
pages = {213--220},
publisher = {North-Holland},
title = {{Non-linear ring currents: effect of strong magnetic fields on $\pi$-electron circulation}},
url = {https://linkinghub.elsevier.com/retrieve/pii/S0009261404017117},
volume = {400},
year = {2004}
}

@article{Tellgren2008,
    title = {{Nonperturbative ab initio calculations in strong magnetic fields using London orbitals}},
    year = {2008},
    journal = {J. Chem. Phys.},
    author = {Tellgren, Erik I. and Soncini, Alessandro and Helgaker, Trygve},
    number = {15},
    month = {10},
    pages = {154114},
    volume = {129},
    url = {http://aip.scitation.org/doi/10.1063/1.2996525},
    doi = {10.1063/1.2996525},
    issn = {0021-9606}
}

@article{Tellgren2014,
author = {Tellgren, E. I. and Teale, A. M. and Furness, J. W. and Lange, K. K. and Ekstr{\"{o}}m, U. and Helgaker, T.},
doi = {10.1063/1.4861427},
issn = {0021-9606},
journal = {J. Chem. Phys.},
month = jan,
number = {3},
pages = {034101},
title = {{Non-perturbative calculation of molecular magnetic properties within current-density functional theory}},
url = {http://aip.scitation.org/doi/10.1063/1.4861427 https://pubs.aip.org/aip/jcp/article/601584},
volume = {140},
year = {2014}
}

@article{Sen2019,
author = {Sen, Sangita and Lange, Kai K. and Tellgren, Erik I.},
doi = {10.1021/acs.jctc.9b00103},
issn = {1549-9618},
journal = {J. Chem. Theory Comput.},
month = {jul},
number = {7},
pages = {3974--3990},
title = {{Excited States of Molecules in Strong Uniform and Nonuniform Magnetic Fields}},
url = {https://pubs.acs.org/doi/10.1021/acs.jctc.9b00103},
volume = {15},
year = {2019}
}

@article{Sun2019b,
author = {Sun, Shichao and Williams-Young, David and Li, Xiaosong},
doi = {10.1021/acs.jctc.9b00095},
issn = {1549-9618},
journal = {J. Chem. Theory Comput.},
month = {may},
number = {5},
pages = {3162--3169},
title = {{An ab Initio Linear Response Method for Computing Magnetic Circular Dichroism Spectra with Nonperturbative Treatment of Magnetic Field}},
url = {https://pubs.acs.org/doi/10.1021/acs.jctc.9b00095},
volume = {15},
year = {2019}
}

@article{Stopkowicz2015,
author = {Stopkowicz, Stella and Gauss, J{\"{u}}rgen and Lange, Kai K. and Tellgren, Erik I. and Helgaker, Trygve},
doi = {10.1063/1.4928056},
issn = {0021-9606},
journal = {J. Chem. Phys.},
month = aug,
number = {7},
pages = {074110},
title = {{Coupled-cluster theory for atoms and molecules in strong magnetic fields}},
url = {http://aip.scitation.org/doi/10.1063/1.4928056},
volume = {143},
year = {2015}
}

@article{Hampe2017,
author = {Hampe, Florian and Stopkowicz, Stella},
doi = {10.1063/1.4979624},
issn = {0021-9606},
journal = {J. Chem. Phys.},
month = apr,
number = {15},
pages = {154105},
pmid = {28433009},
title = {{Equation-of-motion coupled-cluster methods for atoms and molecules in strong magnetic fields}},
url = {http://aip.scitation.org/doi/10.1063/1.4979624},
volume = {146},
year = {2017}
}

@article{Hampe2019,
    title = {{Transition-Dipole Moments for Electronic Excitations in Strong Magnetic Fields Using Equation-of-Motion and Linear Response Coupled-Cluster Theory}},
    year = {2019},
    journal = {J. Chem. Theory Comput.},
    author = {Hampe, Florian and Stopkowicz, Stella},
    number = {7},
    month = {7},
    pages = {4036--4043},
    volume = {15},
    url = {https://pubs.acs.org/doi/10.1021/acs.jctc.9b00242},
    doi = {10.1021/acs.jctc.9b00242},
    issn = {1549-9618}
}

@article{Hampe2020,
author = {Hampe, Florian and Gross, Niklas and Stopkowicz, Stella},
doi = {10.1039/D0CP04169F},
issn = {1463-9076},
journal = {Phys. Chem. Chem. Phys.},
number = {41},
pages = {23522--23529},
title = {{Full triples contribution in coupled-cluster and equation-of-motion coupled-cluster methods for atoms and molecules in strong magnetic fields}},
url = {http://xlink.rsc.org/?DOI=D0CP04169F},
volume = {22},
year = {2020}
}

@article{Furness2015,
author = {Furness, James W. and Verbeke, Joachim and Tellgren, Erik I. and Stopkowicz, Stella and Ekstr{\"{o}}m, Ulf and Helgaker, Trygve and Teale, Andrew M.},
doi = {10.1021/acs.jctc.5b00535},
issn = {1549-9618},
journal = {J. Chem. Theory Comput.},
month = {sep},
number = {9},
pages = {4169--4181},
title = {{Current Density Functional Theory Using Meta-Generalized Gradient Exchange-Correlation Functionals}},
url = {https://pubs.acs.org/doi/10.1021/acs.jctc.5b00535},
volume = {11},
year = {2015}
}

@article{Reimann2019,
author = {Reimann, Sarah and Borgoo, Alex and Austad, Jon and Tellgren, Erik I. and Teale, Andrew M. and Helgaker, Trygve and Stopkowicz, Stella},
doi = {10.1080/00268976.2018.1495849},
issn = {0026-8976},
journal = {Mol. Phys.},
month = {jan},
number = {1},
pages = {97--109},
title = {{Kohn–Sham energy decomposition for molecules in a magnetic field}},
url = {https://www.tandfonline.com/doi/full/10.1080/00268976.2018.1495849},
volume = {117},
year = {2019}
}

@article{Lehtola2020,
archivePrefix = {arXiv},
arxivId = {1812.06274},
author = {Lehtola, Susi and Dimitrova, Maria and Sundholm, Dage},
doi = {10.1080/00268976.2019.1597989},
eprint = {1812.06274},
issn = {0026-8976},
journal = {Mol. Phys.},
month = jan,
number = {2},
pages = {e1597989},
publisher = {Taylor {\&} Francis},
title = {{Fully numerical electronic structure calculations on diatomic molecules in weak to strong magnetic fields}},
url = {https://www.tandfonline.com/doi/abs/10.1080/00268976.2019.1597989 https://www.tandfonline.com/doi/full/10.1080/00268976.2019.1597989 https://doi.org/10.1080/00268976.2019.1597989},
volume = {118},
year = {2020}
}

@article{Pausch2020,
    title = {{Efficient evaluation of three-centre two-electron integrals over London orbitals}},
    year = {2020},
    journal = {Mol. Phys.},
    author = {Pausch, Ansgar and Klopper, Wim},
    number = {21-22},
    month = {11},
    pages = {e1736675},
    volume = {118},
    url = {https://www.tandfonline.com/doi/full/10.1080/00268976.2020.1736675},
    doi = {10.1080/00268976.2020.1736675},
    issn = {0026-8976}
}

@article{Pausch2021,
author = {Pausch, Ansgar and Gebele, Melanie and Klopper, Wim},
doi = {10.1063/5.0069859},
issn = {0021-9606},
journal = {J. Chem. Phys.},
month = nov,
number = {20},
pages = {201101},
pmid = {34852467},
publisher = {AIP Publishing, LLC},
title = {{Molecular point groups and symmetry in external magnetic fields}},
url = {https://doi.org/10.1063/5.0069859 https://aip.scitation.org/doi/10.1063/5.0069859},
volume = {155},
year = {2021}
}

@article{Blaschke2021,
archivePrefix = {arXiv},
arxivId = {2108.11370},
author = {Blaschke, Simon and Stopkowicz, Stella},
doi = {10.1063/5.0076588},
eprint = {2108.11370},
issn = {0021-9606},
journal = {J. Chem. Phys.},
month = jan,
title = {{Cholesky decomposition of complex two-electron integrals over GIAOs: Efficient MP2 computations for large molecules in strong magnetic fields}},
url = {http://arxiv.org/abs/2108.11370 https://aip.scitation.org/doi/10.1063/5.0076588},
year = {2022}
}

@article{Monzel2022,
author = {Monzel, Laurenz and Pausch, Ansgar and Peters, Laurens D. M. and Tellgren, Erik I. and Helgaker, Trygve and Klopper, Wim},
doi = {10.1063/5.0097800},
issn = {0021-9606},
journal = {J. Chem. Phys.},
month = aug,
number = {5},
pages = {054106},
publisher = {AIP Publishing LLC},
title = {{Molecular dynamics of linear molecules in strong magnetic fields}},
url = {https://aip.scitation.org/doi/abs/10.1063/5.0097800 https://pubs.aip.org/aip/jcp/article/2841531},
volume = {157},
year = {2022}
}

@article{Pemberton2022,
author = {Pemberton, Miles J. and Irons, Tom J. P. and Helgaker, Trygve and Teale, Andrew M.},
doi = {10.1063/5.0092520},
issn = {0021-9606},
journal = {J. Chem. Phys.},
month = may,
number = {20},
pages = {204113},
pmid = {35649858},
publisher = {AIP Publishing LLCAIP Publishing},
title = {{Revealing the exotic structure of molecules in strong magnetic fields}},
url = {https://aip.scitation.org/doi/abs/10.1063/5.0092520 https://aip.scitation.org/doi/10.1063/5.0092520},
volume = {156},
year = {2022}
}

@article{Garstang1977,
author = {Garstang, R H},
doi = {10.1088/0034-4885/40/2/001},
issn = {0034-4885},
journal = {Reports Prog. Phys.},
month = feb,
number = {2},
pages = {105--154},
publisher = {IOP Publishing},
title = {{Atoms in high magnetic fields (white dwarfs)}},
url = {https://iopscience.iop.org/article/10.1088/0034-4885/40/2/001 https://iopscience.iop.org/article/10.1088/0034-4885/40/2/001/meta},
volume = {40},
year = {1977}
}

@article{Jordan2008,
author = {Jordan, Stefan},
doi = {10.1017/S1743921309030749},
issn = {1743-9213},
journal = {Proc. Int. Astron. Union},
month = nov,
number = {S259},
pages = {369--378},
title = {{Magnetic fields in White Dwarfs and their direct progenitors}},
url = {https://www.cambridge.org/core/product/identifier/S1743921309030749/type/journal_article},
volume = {4},
year = {2008}
}

@article{Ferrario2015,
author = {Ferrario, Lilia and de Martino, Domitilla and G{\"{a}}nsicke, Boris T.},
doi = {10.1007/s11214-015-0152-0},
issn = {0038-6308},
journal = {Space Sci. Rev.},
month = oct,
number = {1-4},
pages = {111--169},
title = {{Magnetic White Dwarfs}},
url = {http://link.springer.com/10.1007/s11214-015-0152-0},
volume = {191},
year = {2015}
}

@article{Kawka2019,
author = {Kawka, Adela and Vennes, St{\'{e}}phane and Ferrario, Lilia and Paunzen, Ernst},
doi = {10.1093/mnras/sty3048},
issn = {0035-8711},
journal = {Mon. Not. R. Astron. Soc.},
month = feb,
number = {4},
pages = {5201--5210},
title = {{Evidence of enhanced magnetism in cool, polluted white dwarfs}},
url = {https://academic.oup.com/mnras/article/482/4/5201/5185103},
volume = {482},
year = {2019}
}

@article{Wickramasinghe2000,
author = {Wickramasinghe, D. T. and Ferrario, Lilia},
doi = {10.1086/316593},
issn = {0004-6280},
journal = {Publ. Astron. Soc. Pacific},
month = jul,
number = {773},
pages = {873--924},
title = {{Magnetism in Isolated and Binary White Dwarfs}},
url = {http://iopscience.iop.org/article/10.1086/316593},
volume = {112},
year = {2000}
}

@book{Deglinnocenti2004,
address = {Dordrecht},
author = {Degl'Innocenti, Egidio Landi and Landolfi, Marco},
doi = {10.1007/1-4020-2415-0},
isbn = {978-1-4020-2414-6},
publisher = {Springer Netherlands},
title = {{Polarization in Spectral Lines}},
url = {http://link.springer.com/10.1007/1-4020-2415-0},
year = {2004}
}

@article{Berdyugina2007,
author = {Berdyugina, S. V. and Berdyugin, A. V. and Piirola, V.},
doi = {10.1103/PhysRevLett.99.091101},
issn = {0031-9007},
journal = {Phys. Rev. Lett.},
month = aug,
number = {9},
pages = {091101},
title = {{Molecular Magnetic Dichroism in Spectra of White Dwarfs}},
url = {https://link.aps.org/doi/10.1103/PhysRevLett.99.091101},
volume = {99},
year = {2007}
}

@article{Henry1984,
author = {Henry, R. J. W. and Oconnell, R. F.},
doi = {10.1086/184314},
issn = {0004-637X},
journal = {Astrophys. J.},
month = jul,
pages = {L97},
title = {{On the magnetic field in the white dwarf GRW + 70.8247 deg}},
url = {http://adsabs.harvard.edu/doi/10.1086/184314},
volume = {282},
year = {1984}
}

@article{Henry1985,
author = {Henry, R. J. W. and Oconnell, R. F.},
doi = {10.1086/131540},
issn = {0004-6280},
journal = {Publ. Astron. Soc. Pacific},
month = apr,
pages = {333},
title = {{Hydrogen spectrum in magnetic white dwarfs - H-alpha, H-beta and H-gamma transitions}},
url = {http://iopscience.iop.org/article/10.1086/131540},
volume = {97},
year = {1985}
}

@article{Forster1984,
author = {Forster, H and Strupat, W and Rosner, W and Wunner, G and Ruder, H and Herold, H},
doi = {10.1088/0022-3700/17/7/015},
issn = {0022-3700},
journal = {J. Phys. B At. Mol. Phys.},
month = apr,
number = {7},
pages = {1301--1319},
title = {{Hydrogen atoms in arbitrary magnetic fields. II. Bound-bound transitions}},
url = {https://iopscience.iop.org/article/10.1088/0022-3700/17/7/015},
volume = {17},
year = {1984}
}

@article{Greenstein1984,
author = {Greenstein, Jesse L.},
doi = {10.1086/184282},
issn = {0004-637X},
journal = {Astrophys. J.},
month = jun,
pages = {L47},
title = {{The identification of hydrogen in GRW +70 deg 8247}},
url = {http://adsabs.harvard.edu/doi/10.1086/184282},
volume = {281},
year = {1984}
}

@article{Greenstein1985,
author = {Greenstein, J. L. and Henry, R. J. W. and Oconnell, R. F.},
doi = {10.1086/184427},
issn = {0004-637X},
journal = {Astrophys. J.},
month = feb,
pages = {L25},
title = {{Futher identifications of hydrogen in GRW +708247}},
url = {http://adsabs.harvard.edu/doi/10.1086/184427},
volume = {289},
year = {1985}
}

@article{Schmidt1996,
author = {Schmidt, Gary D. and Allen, Richard G. and Smith, Paul S. and Liebert, James},
doi = {10.1086/177244},
issn = {0004-637X},
journal = {Astrophys. J.},
month = may,
pages = {320},
title = {{Combined Ultraviolet-Optical Spectropolarimetry of the Magnetic White Dwarf GD 229}},
url = {http://adsabs.harvard.edu/doi/10.1086/177244},
volume = {463},
year = {1996}
}

@article{Schmidt1996err,
author = {Schmidt, Gary D. and Allen, Richard G. and Smith, Paul S. and Liebert, James},
doi = {10.1086/178169},
issn = {0004-637X},
journal = {Astrophys. J.},
month = dec,
number = {1},
pages = {569--569},
title = {{Erratum: “Combined Ultraviolet‐Optical Spectropolarimetry of the Magnetic White Dwarf GD 229” (ApJ, 463, 320 [1996])}},
url = {https://iopscience.iop.org/article/10.1086/178169},
volume = {473},
year = {1996}
}

@article{Dufour2007,
author = {Dufour, P. and Liebert, J. and Fontaine, G. and Behara, N.},
doi = {10.1038/nature06318},
issn = {0028-0836},
journal = {Nature},
month = nov,
number = {7169},
pages = {522--524},
title = {{White dwarf stars with carbon atmospheres}},
url = {http://www.nature.com/articles/nature06318},
volume = {450},
year = {2007}
}

@article{Dongarra1990,
    title = {{A set of level 3 basic linear algebra subprograms}},
    year = {1990},
    journal = {ACM Transactions on Mathematical Software},
    author = {Dongarra, J. J. and Du Croz, Jeremy and Hammarling, Sven and Duff, I. S.},
    number = {1},
    month = {3},
    pages = {1--17},
    volume = {16},
    url = {https://dl.acm.org/doi/10.1145/77626.79170},
    doi = {10.1145/77626.79170},
    issn = {0098-3500}
}

@article{London1937,
    title = {{Th{\'{e}}orie quantique des courants interatomiques dans les combinaisons aromatiques}},
    year = {1937},
    journal = {J. Phys. Radium},
    author = {London, F.},
    number = {10},
    pages = {397--409},
    volume = {8},
    url = {http://www.edpsciences.org/10.1051/jphysrad:01937008010039700},
    doi = {10.1051/jphysrad:01937008010039700},
    issn = {0368-3842}
}

@MISC{QCUMBRE,
  author = {Hampe, Florian and Stopkowicz, Stella and Groß, Niklas and Kitsaras, Marios-Petros and Grazioli, Laura and Blaschke, Simon and Monzel, Laurenz and Yerg{\"u}n, {\"U}. P.},
  title = {{QCUMBRE}, Quantum Chemical Utility enabling Magnetic-field dependent investigations Benefitting from Rigorous Electron-correlation treatment},
   note = {qcumbre.org}
}

@article{Matthews2020,
    title = {{Coupled-cluster techniques for computational chemistry: The CFOUR program package}},
    year = {2020},
    journal = {J. Chem. Phys.},
    author = {Matthews, Devin A. and Cheng, Lan and Harding, Michael E. and Lipparini, Filippo and Stopkowicz, Stella and Jagau, Thomas-C. and Szalay, Péter G. and Gauss, Jürgen and Stanton, John F.},
    number = {21},
    month = {6},
    pages = {214108},
    volume = {152},
    url = {http://aip.scitation.org/doi/10.1063/5.0004837},
    doi = {10.1063/5.0004837},
    issn = {0021-9606}
}

@misc{cfour,
	Author = {J. F. Stanton and J. Gauss and L. Cheng and M. E. Harding and D. A. Matthews and P. G. Szalay},
	Note = {{W}ith contributions from {A}. {A}sthana, {A}.{A}. {A}uer, {R}.{J}. {B}artlett, {U}. {B}enedikt, {C}. {B}erger,
        {D}.{E}. {B}ernholdt, {S}. {B}laschke, {Y}. {J}. {B}omble, {S}. {B}urger, {O}. {C}hristiansen, {D}. {D}atta, {F}. {E}ngel, {R}. {F}aber, {J}. {G}reiner, {M}. {H}eckert,
        {O}. {H}eun, {M}. Hilgenberg, {C}. {H}uber, {T}.-{C}. {J}agau, {D}. {J}onsson, {J}. {J}us{\'e}lius, 
        {T}. Kirsch, {M}.-{P}. {K}itsaras, {K}. {K}lein, {G}.{M}. {K}opper, {W}.{J}. {L}auderdale, {F}. {L}ipparini, {J}. {L}iu, {T}. {M}etzroth, {L}.{A}. {M}{\"u}ck,
        {D}.{P}. {O}'{N}eill, {T}. {N}ottoli, {J}. {O}swald, {D}.{R}. {P}rice, {E}. {P}rochnow, {C}. {P}uzzarini, {K}. {R}uud, {F}. {S}chiffmann,
        {W}. {S}chwalbach, {C}. {S}immons, {S}. {S}topkowicz, {A}. {T}ajti, {T.} Uhlirova, {J}. {V}{\'a}zquez, {F}. {W}ang, {J}.{D}. {W}atts, {P.} Yerg{\"u}n. {C}. {Z}hang, {X}. {Z}heng,
        and the integral packages {MOLECULE} ({J}. {A}lml{\"o}f and {P}.{R}. {T}aylor), {PROPS} ({P}.{R}. {T}aylor),
        {ABACUS} ({T}. {H}elgaker, {H}.{J}. {A}a. {J}ensen, {P}. {J}{\o}rgensen, and {J}. {O}lsen),
        and {ECP} routines by {A}. {V}. {M}itin and {C}. van {W}{\"u}llen. {F}or the current version, see http://www.cfour.de.},
	Title = {{CFOUR, Coupled-Cluster techniques for Computational Chemistry, a quantum-chemical program package}}}

@MISC{MINT,
  author = {J. Gauss and F. Lipparini and S. Burger and S. Blaschke and M.-P. Kitsaras and T. Nottoli and J. Oswald and S. Stopkowicz},
  year = {2015-2023},
  title = {{MINT, Mainz INTegral  package}},
  note = {{Johannes Gutenberg-Universität Mainz, unpublished}}
}

@MISC{LONDON,
  author = {E. Tellgren and T. Helgaker and A. Soncini and K. K. Lange and A. M. Teale and U. Ekström and S. Stopkowicz and J. H. Austad and S. Sen},
  title = {{LONDON}, a quantum-chemistry program for plane-wave/GTO hybrid basis sets and finite magnetic field calculations},
  note = {londonprogram.org}
}

@misc{quest,
	Title = {{QUEST, A rapid development platform for QUantum Electronic
Structure Techniques, 2017; \\ quest.codes}}}

@misc{bagel,
	Title = {{BAGEL, Brilliantly Advanced General Electronic-structure Library.
http://www.nubakery.org under the GNU General Public License.}}}

@article{Bischoff2020,
author = {Bischoff, Florian A.},
doi = {10.1103/PhysRevA.101.053413},
issn = {2469-9926},
journal = {Phys. Rev. A},
month = may,
number = {5},
pages = {053413},
publisher = {American Physical Society},
title = {{Structure of the H$_3$ molecule in a strong homogeneous magnetic field as computed by the Hartree-Fock method using multiresolution analysis}},
url = {https://journals.aps.org/pra/abstract/10.1103/PhysRevA.101.053413 https://link.aps.org/doi/10.1103/PhysRevA.101.053413},
volume = {101},
year = {2020}
}

@article{Cizek1966,
    title = {{On the Correlation Problem in Atomic and Molecular Systems. Calculation of Wavefunction Components in Ursell‐Type Expansion Using Quantum‐Field Theoretical Methods}},
    year = {1966},
    journal = {J. Chem. Phys.},
    author = {{\v{C}}{\'{i}}{\v{z}}ek, Jiří},
    number = {11},
    month = {12},
    pages = {4256--4266},
    volume = {45},
    url = {http://aip.scitation.org/doi/10.1063/1.1727484},
    doi = {10.1063/1.1727484},
    issn = {0021-9606}
}

@article{Monkhorst1977,
author = {Monkhorst, Hendrik J.},
doi = {10.1002/qua.560120850},
issn = {00207608},
journal = {Int. J. Quantum Chem.},
month = {jan},
number = {S11},
pages = {421--432},
title = {{Calculation of properties with the coupled-cluster method}},
url = {https://onlinelibrary.wiley.com/doi/10.1002/qua.560120850},
volume = {12},
year = {1977}
}

@book{Shavitt2009,
    title = {{Many – Body Methods in Chemistry and Physics}},
    year = {2009},
    booktitle = {Many - Body Methods in Chemistry and Physics: MBPT and Coupled-Cluster Theory},
    author = {Shavitt, Isaiah and Bartlett, Rodney J.},
    publisher = {Cambridge University Press},
    url = {http://ebooks.cambridge.org/ref/id/CBO9780511596834},
    address = {Cambridge},
    isbn = {9780511596834},
    doi = {10.1017/CBO9780511596834}
}

@book{Bishop1973,
author = {Bishop, David M.},
publisher = {Dover Publications},
title = {{Group Theory and Chemistry}},
year = {1973}
}

@article{Davidson1975s,
author = {Davidson, Ernest R.},
doi = {10.1063/1.430484},
issn = {00219606},
journal = {J. Chem. Phys.},
number = {2},
pages = {400},
title = {{Use of double cosets in constructing integrals over symmetry orbitals}},
url = {http://scitation.aip.org/content/aip/journal/jcp/62/2/10.1063/1.430484},
volume = {62},
year = {1975}
}

@article{Taylor1986,
author = {Taylor, Peter R},
doi = {10.1007/BF00526703},
issn = {0040-5744},
journal = {Theor. Chim. Acta},
month = {jun},
number = {5-6},
pages = {447--460},
title = {{Symmetry-adapted integral derivatives}},
url = {http://link.springer.com/10.1007/BF00526703},
volume = {69},
year = {1986}
}

@article{Pausch2021s,
author = {Pausch, Ansgar and Gebele, Melanie and Klopper, Wim},
doi = {10.1063/5.0069859},
issn = {0021-9606},
journal = {J. Chem. Phys.},
month = {nov},
number = {20},
pages = {201101},
title = {{Molecular point groups and symmetry in external magnetic fields}},
url = {https://aip.scitation.org/doi/10.1063/5.0069859},
volume = {155},
year = {2021}
}

@unpublished{unpKitsaras2022,
author = {Kitsaras, Marios-Petros and Stopkowicz, Stella},
title= {Exploiting symmetry in quantum-chemical calculations in finite magnetic field: Abelian complex groups},
note = "{in preparation}",
year={2023}
}

@article{Epifanovsky2013,
author = {Epifanovsky, Evgeny and Zuev, Dmitry and Feng, Xintian and Khistyaev, Kirill and Shao, Yihan and Krylov, Anna I.},
doi = {10.1063/1.4820484},
issn = {0021-9606},
journal = {J. Chem. Phys.},
month = {oct},
number = {13},
pages = {134105},
title = {{General implementation of the resolution-of-the-identity and Cholesky representations of electron repulsion integrals within coupled-cluster and equation-of-motion methods: Theory and benchmarks}},
url = {http://aip.scitation.org/doi/10.1063/1.4820484},
volume = {139},
year = {2013}
}

@article{WilliamsYoung2020,
    title = {{The Chronus Quantum software package}},
    year = {2020},
    journal = {WIREs Computational Molecular Science},
    author = {Williams‐Young, David B. and Petrone, Alessio and Sun, Shichao and Stetina, Torin F. and Lestrange, Patrick and Hoyer, Chad E. and Nascimento, Daniel R. and Koulias, Lauren and Wildman, Andrew and Kasper, Joseph and Goings, Joshua J. and Ding, Feizhi and DePrince, A. Eugene and Valeev, Edward F. and Li, Xiaosong},
    number = {2},
    month = {3},
    volume = {10},
    url = {https://onlinelibrary.wiley.com/doi/10.1002/wcms.1436},
    %doi = {10.1002/wcms.1436},
    issn = {1759-0876}
}

@article{Folkestad2019,
author = {Folkestad, Sarai D. and Kj{\o}nstad, Eirik F. and Koch, Henrik},
doi = {10.1063/1.5083802},
issn = {0021-9606},
journal = {J. Chem. Phys.},
month = {may},
number = {19},
pages = {194112},
title = {{An efficient algorithm for Cholesky decomposition of electron repulsion integrals}},
url = {http://aip.scitation.org/doi/10.1063/1.5083802},
volume = {150},
year = {2019}
}

@article{Gauss2022,
abstract = {A rigorous analysis is carried out concerning the use of Cholesky decomposition (CD) of two-electron integrals in the case of quantum-chemical calculations with finite or perturbative magnetic fiel...},
author = {Gauss, J{\"{u}}rgen and Blaschke, Simon and Burger, Sophia and Nottoli, Tommaso and Lipparini, Filippo and Stopkowicz, Stella},
%doi = {10.1080/00268976.2022.2101562},
issn = {0026-8976},
journal = {Mol. Phys.},
month = jun,
number = {11-12},
publisher = {Taylor {\&} Francis},
title = {{Cholesky decomposition of two-electron integrals in quantum-chemical calculations with perturbative or finite magnetic fields using gauge-including atomic orbitals}},
url = {https://www.tandfonline.com/doi/full/10.1080/00268976.2022.2101562},
volume = {121},
year = {2023}
}

@article{Reynolds2015,
author = {Reynolds, Ryan D. and Shiozaki, Toru},
doi = {10.1039/C4CP04027A},
isbn = {4853.712062},
issn = {1463-9076},
journal = {Phys. Chem. Chem. Phys.},
month = {may},
number = {22},
pages = {14280--14283},
publisher = {The Royal Society of Chemistry},
title = {{Fully relativistic self-consistent field under a magnetic field}},
url = {https://pubs.rsc.org/en/content/articlehtml/2015/cp/c4cp04027a https://pubs.rsc.org/en/content/articlelanding/2015/cp/c4cp04027a http://xlink.rsc.org/?DOI=C4CP04027A},
volume = {17},
year = {2015}
}

@article{Christiansen1995respcc3,
author = {Christiansen, Ove and Koch, Henrik and J{\o}rgensen, Poul},
doi = {10.1063/1.470315},
journal = {J. Chem. Phys.},
month = nov,
number = {17},
pages = {7429--7441},
title = {{Response functions in the CC3 iterative triple excitation model}},
url = {http://aip.scitation.org/doi/10.1063/1.470315},
volume = {103},
year = {1995}
}

@article{Christiansen1995cc2,
author = {Christiansen, Ove and Koch, Henrik and J{\o}rgensen, Poul},
doi = {10.1016/0009-2614(95)00841-Q},
issn = {00092614},
journal = {Chem. Phys. Lett.},
month = sep,
number = {5-6},
pages = {409--418},
title = {{The second-order approximate coupled cluster singles and doubles model CC2}},
url = {https://linkinghub.elsevier.com/retrieve/pii/000926149500841Q},
volume = {243},
year = {1995}
}

@article{Wiebeler2021,
author = {Wiebeler, Christian and Vollbrecht, Joachim and Neuba, Adam and Kitzerow, Heinz-Siegfried and Schumacher, Stefan},
doi = {10.1038/s41598-021-95551-0},
isbn = {0123456789},
issn = {2045-2322},
journal = {Sci. Rep.},
month = {dec},
number = {1},
pages = {16097},
pmid = {34373513},
publisher = {Nature Publishing Group},
title = {{Unraveling the electrochemical and spectroscopic properties of neutral and negatively charged perylene tetraethylesters}},
url = {https://www.nature.com/articles/s41598-021-95551-0},
volume = {11},
year = {2021}
}

@article{Stockett2020,
author = {Stockett, Mark H. and Kj{\ae}r, Christina and Daly, Steven and Bieske, Evan J. and Verlet, Jan R. R. and Nielsen, Steen Br{\o}ndsted and Bull, James N.},
doi = {10.1021/acs.jpca.0c07123},
issn = {1089-5639},
journal = {J. Phys. Chem. A},
month = {oct},
number = {41},
pages = {8429--8438},
title = {{Photophysics of Isolated Rose Bengal Anions}},
url = {https://pubs.acs.org/doi/10.1021/acs.jpca.0c07123},
volume = {124},
year = {2020}
}

@article{Hornum2020,
abstract = {The solvatochromic fluorophore Nile Red, 9-diethylamino-5H-benzo[a]phenoxazine-5-one, is one of the most commonly used stains to enhance contrast of lipid-rich areas of microscopic biosamples. Quite surprisingly, relatively little is known about the spectrally-resolved two-photon absorption (2PA) properties of this dye despite its promising features for two-photon microscopy of biological matter. For this reason, the two-photon solvatochromism of Nile Red still remains an uncharted territory as well. Also, no study has yet reported on how electron-withdrawing substituents attached to the Nile Red backbone affect its solvatochromic properties and two-photon brightness. In this paper, we demonstrate how solvent polarity nfluences the one- and two-photon absorption spectra of Nile Red as well as its fluorescence parameters, and we present new analogues that contain −CF3, −F and −Br substituents on its eastern side. Two-photon excited fluorescence experiments in a broad spectral range (780–1240 nm) and electronic structure calculations show that both the nature and location of the substituent have particular influence on the strength of 2PA, peaking in all cases at approx. 860 and 1050 nm. 2PA cross sections are higher at 1050 nm than at 860 nm, which suggests that Nile Red and its analogues are best suited for two-photon imaging employing excitation in the NIR-II optical transparency window of biological tissues.},
author = {Hornum, Mick and Reinholdt, Peter and Zar{\c{e}} ba, Jan K. and Jensen, Brian B. and W{\"{u}}stner, Daniel and Samo{\'{c}}, Marek and Nielsen, Poul and Kongsted, Jacob},
%doi = {10.1039/d0pp00076k},
file = {:Users/mkitsara/Documents/Mendeley Desktop/One- and two-photon solvatochromism of the fluorescent dye Nile Red and its CF3, F and Br-substituted analogues - Hornum et al. - 2020.pdf:pdf},
issn = {1474-905X},
journal = {Photochem. Photobiol. Sci.},
keywords = {Biochemistry,Biomaterials,Physical Chemistry,Plant Sciences,general},
month = {oct},
number = {10},
pages = {1382--1391},
pmid = {32869822},
publisher = {Springer},
title = {{One- and two-photon solvatochromism of the fluorescent dye Nile Red and its CF3, F and Br-substituted analogues}},
url = {https://link.springer.com/article/10.1039/d0pp00076k https://link.springer.com/10.1039/d0pp00076k},
volume = {19},
year = {2020}
}

@article{Dupuy2019,
author = {Dupuy, Mi-Song and Gloaguen, Eric and Tardivel, Benjamin and Mons, Michel and Brenner, Val{\'{e}}rie},
doi = {10.1021/acs.jctc.9b00923},
issn = {1549-9618},
journal = {J. Chem. Theory Comput.},
month = {jan},
number = {1},
pages = {601--611},
pmid = {31841332},
publisher = {American Chemical Society},
title = {{CC2 Benchmark for Models of Phenylalanine Protein Chains: 0–0 Transition Energies and IR Signatures of the $\pi$$\pi$* Excited State}},
url = {https://pubs.acs.org/doi/full/10.1021/acs.jctc.9b00923 https://pubs.acs.org/doi/10.1021/acs.jctc.9b00923},
volume = {16},
year = {2020}
}

@incollection{Durbeej2020,
  doi = {10.1016/bs.aiq.2020.05.003},
  url = {https://doi.org/10.1016/bs.aiq.2020.05.003},
  year = {2020},
  publisher = {Elsevier},
  pages = {243--268},
  author = {Bo Durbeej},
  title = {Competing excited-state deactivation processes in bacteriophytochromes}
}

@article{Yu2021,
author = {Yu, Xue Fang and Fu, Ting He and Xiao, Bo and Yu, Hong Yuan and Li, Qingzhong},
doi = {10.1039/D1CP00030F},
issn = {1463-9084},
journal = {Phys. Chem. Chem. Phys.},
month = {aug},
number = {30},
pages = {16089--16106},
pmid = {34291779},
publisher = {The Royal Society of Chemistry},
title = {{A theoretical study on the excited-state deactivation paths for the A–5FU dimer}},
url = {https://pubs.rsc.org/en/content/articlehtml/2021/cp/d1cp00030f https://pubs.rsc.org/en/content/articlelanding/2021/cp/d1cp00030f},
volume = {23},
year = {2021}
}

@article{Reinholdt2021,
author = {Reinholdt, Peter and Vidal, Marta L. and Kongsted, Jacob and Iannuzzi, Marcella and Coriani, Sonia and Odelius, Michael},
doi = {10.1021/ACS.JPCLETT.1C02031/ASSET/IMAGES/LARGE/JZ1C02031_0005.JPEG},
issn = {19487185},
journal = {J. Phys. Chem. Lett.},
month = {sep},
number = {36},
pages = {8865--8871},
pmid = {34498464},
publisher = {American Chemical Society},
title = {{NitrogenK-Edge X-ray Absorption Spectra of Ammonium and Ammonia in Water Solution: Assessing the Performance of Polarizable Embedding Coupled Cluster Methods}},
url = {https://pubs.acs.org/doi/full/10.1021/acs.jpclett.1c02031},
volume = {12},
year = {2021}
}

@article{Safin2021,
author = {Safin, Damir A. and Babashkina, Maria G. and Bolte, Michael and Ptaszek, Aleksandra L. and Kuku{\l}ka, Mercedes and Mitoraj, Mariusz P.},
doi = {10.1016/j.jlumin.2021.118264},
issn = {00222313},
journal = {J. Lumin.},
month = {oct},
pages = {118264},
publisher = {North-Holland},
title = {{Novel sterically demanding Schiff base dyes: An insight from experimental and theoretical calculations}},
url = {https://linkinghub.elsevier.com/retrieve/pii/S0022231321003811},
volume = {238},
year = {2021}
}

@article{Kochman2021,
author = {Kochman, Micha{\l} Andrzej and Durbeej, Bo and Kubas, Adam},
doi = {10.1021/ACS.JPCA.1C06166/SUPPL_FILE/JP1C06166_SI_002.ZIP},
issn = {15205215},
journal = {J. Phys. Chem. A},
month = {oct},
number = {39},
pages = {8635--8648},
pmid = {34550700},
publisher = {American Chemical Society},
title = {{Simulation and Analysis of the Transient Absorption Spectrum of 4-(N,N-Dimethylamino)benzonitrile (DMABN) in Acetonitrile}},
url = {https://pubs.acs.org/doi/full/10.1021/acs.jpca.1c06166},
volume = {125},
year = {2021}
}

@article{Naim2021,
author = {Naim, Carmelo and Castet, Fr{\'{e}}d{\'{e}}ric and Matito, Eduard},
doi = {10.1039/D1CP02500G},
issn = {1463-9084},
journal = {Phys. Chem. Chem. Phys.},
month = {sep},
number = {37},
pages = {21227--21239},
publisher = {The Royal Society of Chemistry},
title = {{Impact of van der Waals interactions on the structural and nonlinear optical properties of azobenzene switches}},
url = {https://pubs.rsc.org/en/content/articlehtml/2021/cp/d1cp02500g https://pubs.rsc.org/en/content/articlelanding/2021/cp/d1cp02500g},
volume = {23},
year = {2021}
}

@article{Izsak2020,
abstract = {While methodological developments in the last decade made it possible to compute coupled cluster (CC) energies including excitations up to a perturbative triples correction for molecules containing several hundred atoms, a similar breakthrough has not yet been reported for excited state computations. Accurate CC methods for excited states are still expensive, although some promising candidates for an efficient and accurate excited state CC method have emerged recently. This review examines the various approximation schemes with particular emphasis on their performance for excitation energies and summarizes the best state‐of‐the‐art results which may pave the way for a robust excited state method applicable to molecules of hundreds of atoms. Among these, special attention will be given to exploiting the techniques of similarity transformation, perturbative approximations as well as integral decomposition, local and embedding techniques within the equation of motion CC framework.},
author = {Izs{\'{a}}k, R{\'{o}}bert},
%doi = {10.1002/wcms.1445},
issn = {1759-0876},
journal = {WIREs Comput. Mol. Sci.},
month = {may},
number = {3},
title = {{Single‐reference coupled cluster methods for computing excitation energies in large molecules: The efficiency and accuracy of approximations}},
url = {https://onlinelibrary.wiley.com/doi/10.1002/wcms.1445 https://wires.onlinelibrary.wiley.com/doi/10.1002/wcms.1445},
volume = {10},
year = {2020}
}

@incollection{Hattig2006,
author = {H{\"{a}}ttig, Christof},
booktitle = {Comput. Nanosci. Do It Yours.},
editor = {Grotendorst, J. and Bl{\"{u}}gel, S. and Marx, D},
isbn = {3000173501},
pages = {245--278},
publisher = {John von Neumann Institute for Computing},
title = {{Beyond Hartree-Fock: MP2 and Coupled-Cluster Methods for Large Systems}},
url = {http://www.fz-juelich.de/nic-series/volume31},
volume = {31},
year = {2006}
}

@article{Schreiber2008,
author = {Schreiber, Marko and Silva-Junior, Mario R. and Sauer, Stephan P. A. and Thiel, Walter},
doi = {10.1063/1.2889385},
issn = {0021-9606},
journal = {J. Chem. Phys.},
month = apr,
number = {13},
pages = {134110},
title = {{Benchmarks for electronically excited states: CASPT2, CC2, CCSD, and CC3}},
url = {http://aip.scitation.org/doi/10.1063/1.2889385},
volume = {128},
year = {2008}
}

@article{Purvis1982,
author = {Purvis, George D. and Bartlett, Rodney J.},
doi = {10.1063/1.443164},
issn = {0021-9606},
journal = {J. Chem. Phys.},
month = {feb},
number = {4},
pages = {1910--1918},
title = {{A full coupled‐cluster singles and doubles model: The inclusion of disconnected triples}},
url = {http://aip.scitation.org/doi/10.1063/1.443164},
volume = {76},
year = {1982}
}

@incollection{Krylov2017,
author = {Krylov, Anna I.},
booktitle = {Rev. Comput. Chem.},
doi = {10.1002/9781119356059.ch4},
editor = {Parrill, Abby L. and Lipkowitz, Kenny B.},
month = {apr},
pages = {151--224},
publisher = {John Wiley {\&} Sons, Ltd},
title = {{The Quantum Chemistry of Open-Shell Species}},
url = {http://doi.wiley.com/10.1002/9781119356059.ch4 https://onlinelibrary.wiley.com/doi/10.1002/9781119356059.ch4},
year = {2017}
}

@article{Christiansen1995cc3,
author = {Christiansen, Ove and Koch, Henrik and J{\o}rgensen, Poul},
doi = {10.1063/1.470315},
issn = {0021-9606},
journal = {J. Chem. Phys.},
month = {nov},
number = {17},
pages = {7429--7441},
title = {{Response functions in the CC3 iterative triple excitation model}},
url = {http://aip.scitation.org/doi/10.1063/1.470315},
volume = {103},
year = {1995}
}

@article{Koch1997,
author = {Koch, Henrik and Christiansen, Ove and J{\o}rgensen, Poul and {Sanchez de Mer{\'{a}}s}, Alfredo M. and Helgaker, Trygve},
doi = {10.1063/1.473322},
issn = {0021-9606},
journal = {J. Chem. Phys.},
month = {2},
number = {5},
pages = {1808--1818},
title = {{The CC3 model: An iterative coupled cluster approach including connected triples}},
url = {http://aip.scitation.org/doi/10.1063/1.473322},
volume = {106},
year = {1997}
}

@article{Bondanza2021,
author = {Bondanza, Mattia and Jacquemin, Denis and Mennucci, Benedetta},
doi = {10.1021/acs.jpclett.1c01929},
issn = {1948-7185},
journal = {J. Phys. Chem. Lett.},
month = {jul},
number = {28},
pages = {6604--6612},
title = {{Excited States of Xanthophylls Revisited: Toward the Simulation of Biologically Relevant Systems}},
url = {https://pubs.acs.org/doi/10.1021/acs.jpclett.1c01929},
volume = {12},
year = {2021}
}

@article{Davis2021,
author = {Davis, Megan C. and Fortenberry, Ryan C.},
doi = {10.1021/acs.jctc.1c00307},
issn = {1549-9618},
journal = {J. Chem. Theory Comput.},
month = {jul},
number = {7},
pages = {4374--4382},
title = {{(T)+EOM Quartic Force Fields for Theoretical Vibrational Spectroscopy of Electronically Excited States}},
url = {https://pubs.acs.org/doi/10.1021/acs.jctc.1c00307},
volume = {17},
year = {2021}
}

@article{Bilalbegovic2021,
author = {Bilalbegovi{\'{c}}, Goranka and Maksimovi{\'{c}}, Aleksandar and Valencic, Lynne A. and Lehtola, Susi},
doi = {10.1021/acsearthspacechem.0c00238},
issn = {2472-3452},
journal = {ACS Earth Space Chem.},
month = {mar},
number = {3},
pages = {436--448},
title = {{Sulfur Molecules in Space by X-rays: A Computational Study}},
url = {https://pubs.acs.org/doi/10.1021/acsearthspacechem.0c00238},
volume = {5},
year = {2021}
}

@article{Hutcheson2021,
author = {Hutcheson, Anders and Paul, Alexander Christian and Myhre, Rolf H. and Koch, Henrik and H{\o}yvik, Ida‐Marie},
doi = {10.1002/jcc.26553},
issn = {0192-8651},
journal = {J. Comput. Chem.},
month = {jul},
number = {20},
pages = {1419--1429},
title = {{Describing ground and excited state potential energy surfaces for molecular photoswitches using coupled cluster models}},
url = {https://onlinelibrary.wiley.com/doi/10.1002/jcc.26553},
volume = {42},
year = {2021}
}

@article{Neville2018,
author = {Neville, Simon P. and Stolow, Albert and Schuurman, Michael S.},
doi = {10.1063/1.5044392},
issn = {0021-9606},
journal = {J. Chem. Phys.},
month = {oct},
number = {14},
pages = {144310},
title = {{Vacuum ultraviolet excited state dynamics of the smallest ring, cyclopropane. I. A reinterpretation of the electronic spectrum and the effect of intensity borrowing}},
url = {http://aip.scitation.org/doi/10.1063/1.5044392},
volume = {149},
year = {2018}
}

@article{Fedotov2021,
author = {Fedotov, Daniil A. and Paul, Alexander C. and Posocco, Paolo and Santoro, Fabrizio and Garavelli, Marco and Koch, Henrik and Coriani, Sonia and Improta, Roberto},
doi = {10.1021/acs.jctc.0c01150},
issn = {1549-9618},
journal = {J. Chem. Theory Comput.},
month = {mar},
number = {3},
pages = {1638--1652},
title = {{Excited-State Absorption of Uracil in the Gas Phase: Mapping the Main Decay Paths by Different Electronic Structure Methods}},
url = {https://pubs.acs.org/doi/10.1021/acs.jctc.0c01150},
volume = {17},
year = {2021}
}

@article{Veril2021,
abstract = {We describe our efforts of the past few years to create a large set of more than 500 highly accurate vertical excitation energies of various natures ( $\pi$ → $\pi$ * , n → $\pi$ * , double excitation, Rydberg, singlet, doublet, triplet, etc.) in small‐ and medium‐sized molecules. These values have been obtained using an incremental strategy which consists in combining high‐order coupled cluster and selected configuration interaction calculations using increasingly large diffuse basis sets in order to reach high accuracy. One of the key aspects of the so‐called QUEST database of vertical excitations is that it does not rely on any experimental values, avoiding potential biases inherently linked to experiments and facilitating theoretical cross comparisons. Following this composite protocol, we have been able to produce theoretical best estimates (TBEs) with the aug‐cc‐pVTZ basis set for each of these transitions, as well as basis set corrected TBEs (i.e., near the complete basis set limit) for some of them. The TBEs/aug‐cc‐pVTZ have been employed to benchmark a large number of (lower‐order) wave function methods such as CIS(D), ADC(2), CC2, STEOM‐CCSD, CCSD, CCSDR(3), CCSDT‐3, ADC(3), CC3, NEVPT2, and so on (including spin‐scaled variants). In order to gather the huge amount of data produced during the QUEST project, we have created a website ( https://lcpq.github.io/QUESTDB{\_}website ) where one can easily test and compare the accuracy of a given method with respect to various variables such as the molecule size or its family, the nature of the excited states, the type of basis set, and so on. We hope that the present review will provide a useful summary of our effort so far and foster new developments around excited‐state methods.},
author = {V{\'{e}}ril, Micka{\"{e}}l and Scemama, Anthony and Caffarel, Michel and Lipparini, Filippo and Boggio‐Pasqua, Martial and Jacquemin, Denis and Loos, Pierre‐Fran{\c{c}}ois},
%doi = {10.1002/wcms.1517},
issn = {1759-0876},
journal = {WIREs Comput. Mol. Sci.},
month = {sep},
number = {5},
title = {{QUESTDB : A database of highly accurate excitation energies for the electronic structure community}},
url = {https://onlinelibrary.wiley.com/doi/10.1002/wcms.1517 https://wires.onlinelibrary.wiley.com/doi/10.1002/wcms.1517},
volume = {11},
year = {2021}
}

@article{Matthews2020a,
author = {Matthews, Devin A.},
doi = {10.1080/00268976.2020.1771448},
issn = {0026-8976},
journal = {Mol. Phys.},
month = {nov},
number = {21-22},
pages = {e1771448},
title = {{EOM-CC methods with approximate triple excitations applied to core excitation and ionisation energies}},
url = {https://www.tandfonline.com/doi/full/10.1080/00268976.2020.1771448},
volume = {118},
year = {2020}
}

@article{Loos2020,
author = {Loos, Pierre-Fran{\c{c}}ois and Lipparini, Filippo and Boggio-Pasqua, Martial and Scemama, Anthony and Jacquemin, Denis},
doi = {10.1021/acs.jctc.9b01216},
issn = {1549-9618},
journal = {J. Chem. Theory Comput.},
month = {mar},
number = {3},
pages = {1711--1741},
title = {{A Mountaineering Strategy to Excited States: Highly Accurate Energies and Benchmarks for Medium Sized Molecules}},
url = {https://pubs.acs.org/doi/10.1021/acs.jctc.9b01216},
volume = {16},
year = {2020}
}

@article{Moller1934,
author = {M{\o}ller, Chr. and Plesset, M. S.},
doi = {10.1103/PhysRev.46.618},
issn = {0031-899X},
journal = {Phys. Rev.},
month = {oct},
number = {7},
pages = {618--622},
title = {{Note on an Approximation Treatment for Many-Electron Systems}},
url = {https://link.aps.org/doi/10.1103/PhysRev.46.618},
volume = {46},
year = {1934}
}

@article{Paul2021,
archivePrefix = {arXiv},
arxivId = {2007.01088},
author = {Paul, Alexander C. and Myhre, Rolf H. and Koch, Henrik},
doi = {10.1021/acs.jctc.0c00686},
eprint = {2007.01088},
issn = {1549-9618},
journal = {J. Chem. Theory Comput.},
month = jan,
number = {1},
pages = {117--126},
pmid = {33263255},
title = {{New and Efficient Implementation of CC3}},
url = {https://pubs.acs.org/doi/10.1021/acs.jctc.0c00686},
volume = {17},
year = {2021}
}

@article{Wuellen2016,
   author = {Benjamin Helmich-Paris and Christof Hättig and Christoph Van Wüllen},
   doi = {10.1021/ACS.JCTC.5B01197/ASSET/IMAGES/CT-2015-01197H_M070.GIF},
   issn = {15499626},
   issue = {4},
   journal = {J. Chem. Theory Comput.},
   month = {4},
   pages = {1892-1904},
   publisher = {American Chemical Society},
   title = {Spin-Free CC2 Implementation of Induced Transitions between Singlet Ground and Triplet Excited States},
   volume = {12},
   url = {https://pubs.acs.org/doi/full/10.1021/acs.jctc.5b01197},
   year = {2016},
}

@article{Kitsaras2021,
author = {Kitsaras, Marios-Petros and Stopkowicz, Stella},
doi = {10.1063/5.0044362},
issn = {0021-9606},
journal = {J. Chem. Phys.},
month = apr,
number = {13},
pages = {131101},
title = {{Spin contamination in MP2 and CC2, a surprising issue}},
url = {https://aip.scitation.org/doi/10.1063/5.0044362},
volume = {154},
year = {2021}
}

@article{Zuckerman2003,
author = {Zuckerman, B. and Koester, D. and Reid, I. N. and Hunsch, M.},
doi = {10.1086/377492},
issn = {0004-637X},
journal = {Astrophys. J.},
month = oct,
number = {1},
pages = {477--495},
title = {{Metal Lines in DA White Dwarfs}},
url = {https://iopscience.iop.org/article/10.1086/377492},
volume = {596},
year = {2003}
}

@article{Zuckerman2010,
author = {Zuckerman, B. and Melis, C. and Klein, B. and Koester, D. and Jura, M.},
doi = {10.1088/0004-637X/722/1/725},
issn = {0004-637X},
journal = {Astrophys. J.},
month = oct,
number = {1},
pages = {725--736},
title = {ANCIENT PLANETARY SYSTEMS ARE ORBITING A LARGE FRACTION OF WHITE DWARF STARS},
url = {https://iopscience.iop.org/article/10.1088/0004-637X/722/1/725},
volume = {722},
year = {2010}
}

@Misc{nist,
author = {A.~Kramida and {Yu.~Ralchenko} and
J.~Reader and { NIST ASD Team}},
HOWPUBLISHED = {{NIST Atomic Spectra Database
{{https://physics.nist.gov/asd}}.
National Institute of Standards and Technology,
Gaithersburg, MD.}},
year = {2022},
}

@article{Hollands2023,
  doi = {10.1093/mnras/stad143},
  url = {https://doi.org/10.1093/mnras/stad143},
  year = {2023},
  month = jan,
  publisher = {Oxford University Press ({OUP})},
  volume = {520},
  number = {3},
  pages = {3560--3575},
  author = {M A Hollands and S Stopkowicz and M-P Kitsaras and F Hampe and S Blaschke and J J Hermes},
  title = {A {DZ} white dwarf with a 30{\hspace{0.167em}}{MG} magnetic field},
  journal = {Monthly Notices of the Royal Astronomical Society}
}

@article{Lehtola2020a,
archivePrefix = {arXiv},
arxivId = {1812.06274},
author = {Lehtola, Susi and Dimitrova, Maria and Sundholm, Dage},
doi = {10.1080/00268976.2019.1597989},
eprint = {1812.06274},
issn = {0026-8976},
journal = {Mol. Phys.},
month = jan,
number = {2},
pages = {e1597989},
publisher = {Taylor {\&} Francis},
title = {{Fully numerical electronic structure calculations on diatomic molecules in weak to strong magnetic fields}},
url = {https://www.tandfonline.com/doi/abs/10.1080/00268976.2019.1597989 https://www.tandfonline.com/doi/full/10.1080/00268976.2019.1597989 https://doi.org/10.1080/00268976.2019.1597989},
volume = {118},
year = {2020}
}

@phdthesis{Kitsaras2023,
  author       = {Kitsaras, Marios-Petros}, 
  title        = {{Finite magnetic-field Coupled-Cluster methods: Efficiency and Utilities}},
  school       = {{Johannes-Gutenberg Universität Mainz}},
  year         = 2023,
}

@article{pritchard2019a,
    author = {Pritchard, Benjamin P. and Altarawy, Doaa and Didier, Brett and Gibsom, Tara D. and Windus, Theresa L.},
    title = {A New Basis Set Exchange: An Open, Up-to-date Resource for the Molecular Sciences Community},
    journal = {J. Chem. Inf. Model.},
    volume = {59},
    pages = {4814-4820},
    year = {2019},
    doi = {10.1021/acs.jcim.9b00725}
}

@article{feller1996a,
    author = {Feller, David},
    title = {The role of databases in support of computational chemistry calculations},
    journal = {J. Comput. Chem.},
    volume = {17},
    pages = {1571-1586},
    year = {1996},
    doi = {10.1002/(SICI)1096-987X(199610)17:13<1571::AID-JCC9>3.0.CO;2-P}
}

@article{prascher2011a,
    author = {Prascher, Brian P. and Woon, David E. and Peterson, Kirk A. and Dunning, Thom H. and Wilson, Angela K.},
    title = {Gaussian basis sets for use in correlated molecular calculations. VII. Valence, core-valence, and scalar relativistic basis sets for Li, Be, Na, and Mg},
    journal = {Theor. Chem. Acc.},
    volume = {128},
    pages = {69-82},
    year = {2011},
    doi = {10.1007/s00214-010-0764-0}
}

@article{schuchardt2007a,
    author = {Schuchardt, Karen L. and Didier, Brett T. and Elsethagen, Todd and Sun, Lisong and Gurumoorthi, Vidhya and Chase, Jared and Li, Jun and Windus, Theresa L.},
    title = {Basis Set Exchange: A Community Database for Computational Sciences},
    journal = {J. Chem. Inf. Model.},
    volume = {47},
    pages = {1045-1052},
    year = {2007},
    doi = {10.1021/ci600510j}
}

\end{document}